\documentclass{cta-author}

\pdfoutput=1 

\newcommand{\beqn}{\begin{eqnarray}}
\newcommand{\eeqn}{\end{eqnarray}}
\newcommand{\bse}{\begin{subequations}}
\newcommand{\ese}{\end{subequations}}
\newcommand{\D}{\mathrm{d}}
\usepackage{graphicx,color}
\usepackage{braket}
\usepackage{bbm}
\usepackage{url}
\usepackage{soul}
\usepackage{nicematrix}
\usepackage{cases}
\usepackage{subcaption}
\usepackage{enumitem}

{}
{}
{}
\usepackage{comment}
\begin{document}

\supertitle{IET Energy Conversion and Economics}

\title{Solving Differential-Algebraic Equations in Power System Dynamic Analysis with Quantum Computing}

\author{\au{Huynh T. T. Tran$^{1}$}, \au{Hieu T. Nguyen$^{1}$}, \au{Long T. Vu$^{2}$}, \au{Samuel T. Ojetola$^{3}$}}

\address{\add{1}{Department of Electrical \& Computer Engineering, North Carolina Agricultural and Technical State University, Greensboro, NC, 27411, USA}
\add{2}{The Energy \& Environment Directorate, Pacific Northwest National Laboratories, Richland, WA 99354, USA}
\add{3}{The Electric Power System Research, 
Sandia National Laboratories, Albuquerque, NM 87123, USA}
\email{htran@aggies.ncat.edu,  htnguyen1@ncat.edu, thanhlong.vu@pnnl.gov, sojetol@sandia.gov }}

\begin{abstract}
Power system dynamics are generally modeled by high dimensional nonlinear differential-algebraic equations (DAEs) given a large number of components forming the network. 
These DAEs' complexity can grow exponentially due to the increasing penetration of distributed energy resources, whereas their computation time becomes sensitive due to the increasing interconnection of the power grid with other energy systems. 
This paper demonstrates the use of quantum computing algorithms to solve DAEs for power system dynamic analysis.
We leverage a symbolic programming framework to equivalently convert the power system's DAEs into ordinary differential equations (ODEs) using index reduction methods and then encode their data into qubits using amplitude encoding.
The system nonlinearity is captured by Hamiltonian simulation with truncated Taylor expansion so that state variables can be updated by a quantum linear equation solver. 
Our results show that quantum computing can solve the power system's DAEs accurately with a computational complexity polynomial in the logarithm of the system dimension.
We also illustrate the use of recent advanced tools in scientific machine learning for implementing complex computing concepts, i.e. Taylor expansion, DAEs/ODEs transformation, and quantum computing solver with abstract representation for power engineering applications.
\end{abstract}

\maketitle

\section{INTRODUCTION}
\label{intro}

Solving differential-algebraic equations (DAEs) is a fundamental task for time-domain simulation in the power system dynamic analysis where fast computation time and accurate solutions are required \cite{zhou2022quantum}. 
These DAEs include a set of ordinary differential equations (ODEs) modeling the dynamics of synchronous generators, exciters, and governors, along with nonlinear algebraic equations modeling network power flows and Kirchhoff voltage laws for individual buses. 
The scale of these DAEs can be very large due to a large number of grid components such as generators, loads, and transmission lines forming the network and growing exponentially with the increasing penetration of distributed energy resources \cite{sauer2017power}.

Traditionally, we can tackle the power system's DAEs by solving their ODEs using a numerical integration method and solving algebraic network equations by a numerical iteration method at each integration step \cite{sauer2017power,kundur2022power, milano2015current}. 
However, this requires several iterations at each integration step for the convergence of the network equations and a large number of integration steps to guarantee the numerical stability of the ODE solution, leading to huge computation burdens. 
Indeed, numerical methods of DAEs are not as mature as the methods for solving ODEs  \cite{kyriienko2021solving, hairer1993solving}.
Thus, one approach is to convert the DAEs into ODEs for utilizing ODE solution methods \cite{ascher1998computer}.
However, these ODE numerical methods scale exponentially with problem size in classical computers, particularly the number of state variables and algebraic variables, leading to high computational complexity  \cite{stiasny2023solving}.

Quantum computers can achieve algorithmically superior scaling for certain problems, particularly complex linear algebra and matrix exponential \cite{lapierre2021introduction}.
An application with proven quantum advantage is solving linear equations \cite{huang2021near,subacsi2019quantum,lubasch2020variational, harrow2009quantum}. 
The most well-known quantum linear equation solver is the Harrow-Hassidim-Lloyd (HHL) algorithm \cite{harrow2009quantum}, which provides an exponential memory advantage over the classical computing method
i.e., requiring only $\log_2(N)$ qubits for $N$ variables \cite{kyriienko2021solving}. 
Extending quantum linear algebraic equation solvers for high-dimensional linear ODE systems are studied in \cite{berry2014high, childs2017quantum, xin2020quantum}.
By setting the quantum states proportional to the solution of the block-encoded $N$-dimensional system of linear equations, they unlock the potential of rapidly characterizing the solutions of high-dimensional linear ODEs.
However, they cannot tackle DAEs in power system dynamics analysis that contain both nonlinear algebraic equations and nonlinear ODEs.

This paper aims to solve DAEs in power system dynamics using quantum algorithms and recent advances in scientific computing frameworks. 
We leverage symbolic programming packages from Julia/SciML \cite{rackauckas2017differentialequations}, particularly ModelingToolkit \cite{ma2021modelingtoolkit}, to implement and transform the DAEs modeling power system dynamics into equivalent ODEs using the Pantelides based index reduction \cite{Panthe}, which is then tackled by the Leyton-Osborne's quantum algorithm  \cite{leyton2008quantum}. 
Specifically, we employ the second-order Taylor expansion to approximate nonlinear ODEs as polynomial functions of state variables.
These polynomial functions can be encoded in the amplitudes of the tensor of the quantum state and its copy, which in turn can be simulated in quantum computers through Hamiltonian simulation \cite{berry2007efficient}. 
The Forward Euler update of state variables can be then considered as a set of linear equations that can be computed by the quantum linear equation solver, i.e. the HHL algorithm \cite{harrow2009quantum}.
The complexity, remarkably, is polynomial to the logarithm of the system dimension. Within this context, the contributions of our paper are as follows: 
\vspace{-\topsep}
\begin{itemize}[leftmargin = 15pt]
    \item We demonstrate how to convert the computation of solving DAEs in traditional computers into equivalent steps with quantum computing operators, particularly amplitude encoding of data representation, Hamiltonian simulation of modeling nonlinear state functions, and quantum linear solver-based HHL algorithm for updating state variables.

   \item We introduce the use of scientific machine learning, particularly symbolic programming, to facilitate mathematical transformation, such as converting a DAEs system to an ODEs system to ease its computation under quantum computing.
    
\end{itemize}

\vspace{-\topsep}

The rest of the paper is as follows. 
Section \ref{model} presents the mathematical model of power systems dynamics and the classical numerical methods for solving them.
Section \ref{Quantum} provides quantum computing fundamental backgrounds, such as qubits, vector representation, and the time evolution of the quantum state, and discusses the quantum computing algorithm solving power system's DAEs.
Section \ref{implement_and_results} presents our implementation and numerical studies conducted on the single-machine infinite bus system and the three-machine nine-bus test system.
Section \ref{conclu} concludes the paper.

\begin{figure}[t!]
    \centering
    \includegraphics[width = 0.38\textwidth]{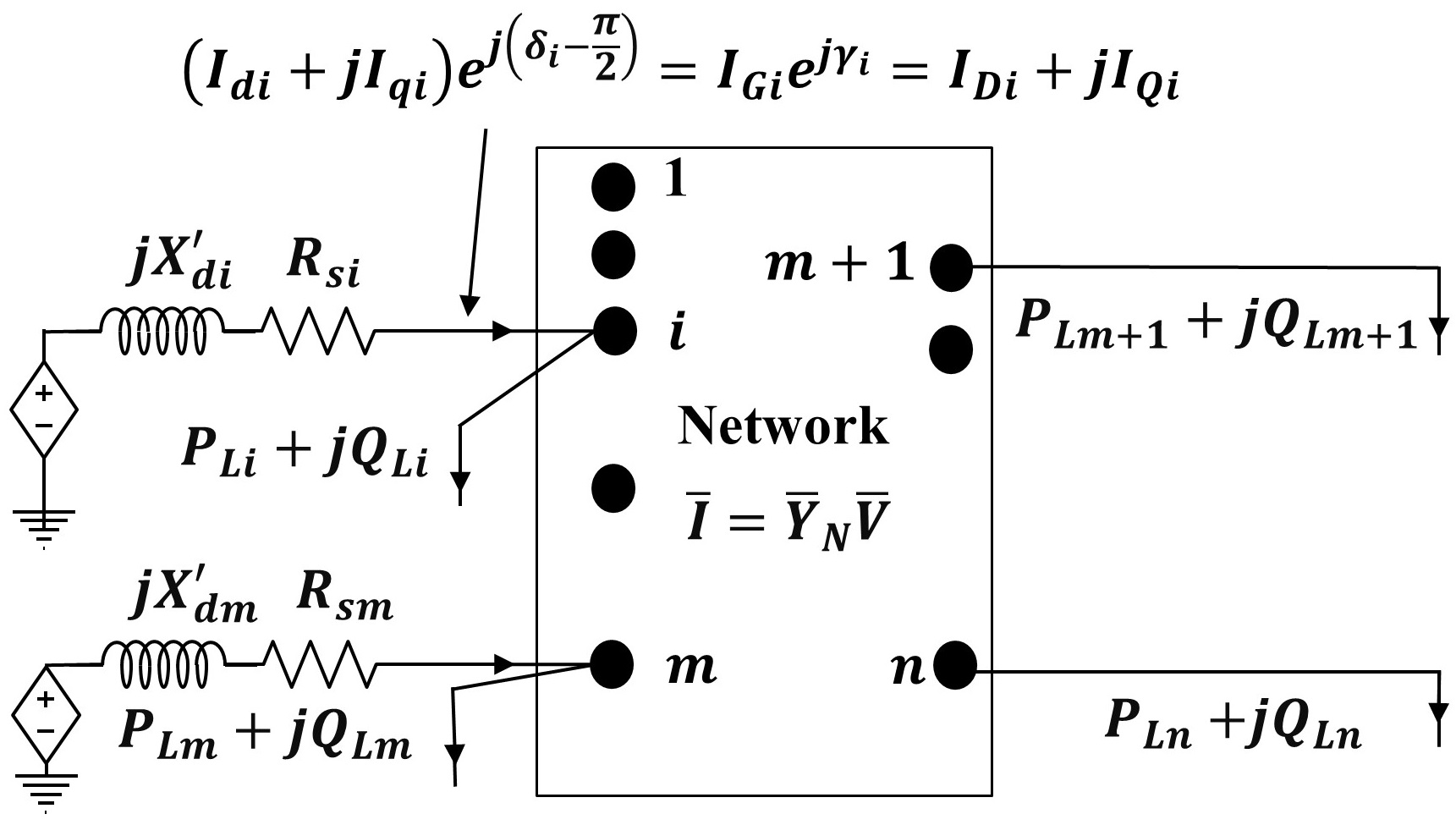} 
    \vspace{-0.5em}
    \caption{Interconnection of $m$ synchronous machine dynamic circuits and $n$ buses of the network \cite{sauer2017power} (page 137).}
    \label{fig:network}
\end{figure}

\section{MATHEMATICAL MODEL}
\label{model}

A generic electric power network can be represented by a dynamic network consisting of $n$ buses with $m$ generators, the power transmission network, and the loads as shown in Figure \ref{fig:network} 
where $\bar{Y}_N$ is the network admitance matrix ($Y_{ik}\angle{\alpha_{ik}}$ represents the $\text{ik}^\text{th}$ element of $\bar{Y}_N$).
Active and reactive power demands at buses $i = 1, \ldots n$ are denoted as $P_{Li}$ and $Q_{Li}$, respectively. 
$V_i$ are the magnitudes and $\theta_i$ is the phase angles of the nodal voltage in bus $i$.
Generation buses are indexed from $1$ to $m$, where the generator in the bus $i$ is represented by a constant voltage source behind the impedance $(R_{si} + jX'_{di})$ ($R_{si}$ is the stator resistance and $X'_{di}$ is the transient reactance).
Here, $I_{di}$, and $I_{qi}$ are the stator currents in $d-q$ coordinates.   
Buses indexed $m+1, \ldots, n$ represent load buses. 
Consequently, the dynamics of the electric power grid generally contain (i) a set of ODEs modeling the system dynamics of synchronous generators along with their equipped governors and stabilizers and (ii) a set of algebraic equations modeling generators' stator voltage equations and power flow balances in the network.
These equations jointly form the DAEs modeling the power system dynamics. 

\subsection{Generic DAE model}

\subsubsection{Dynamics of synchronous generators}
\label{gen_model}
\begin{figure}[t!]
    \centering
    \includegraphics[width = 0.38\textwidth]{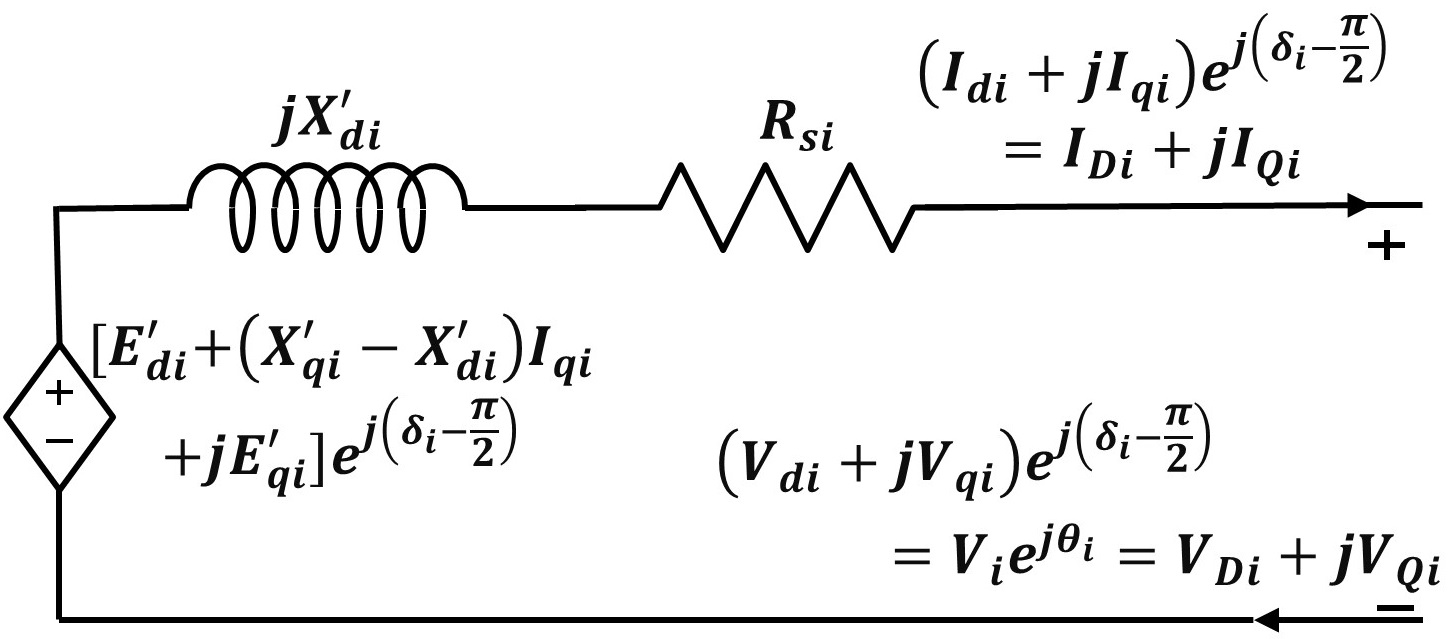}
    \caption{Synchronous machine two-axis model dynamic circuit $(i = 1,\ldots,m)$ \cite{sauer2017power} (page 136).}
    \label{fig:generator}
\end{figure}

We consider the fourth-order two-axis model of synchronous machines represented by a dynamics circuit in the $d-q$ coordinates as shown in Figure \ref{fig:generator} \cite{sauer2017power}.
Its dynamics are as follows:
\bse
\label{gen:dynamics_gen}
\begin{align}
    T'_{qoi}\frac{\D E'_{di}}{\D t} &= -E'_{di} + (X_{qi} - X'_{qi})I_{qi} \label{eq: dynamics_gen_a}  \\
  T'_{doi}\frac{\D E'_{qi}}{\D t} &= -E'_{qi} - (X_{di} - X'_{di})I_{di} + E_{fdi}  \label{eq: dynamics_gen_b} \\
    \frac{\D \delta_{i}}{\D t} & = \omega_{i} - \omega_{s}  \label{eq: dynamics_gen_c} \\
    \frac{2C_i}{\omega_{s}} \frac{\D \omega_{i}}{\D t} & = T_{Mi} - E'_{di}I_{di} -  E'_{qi}I_{qi} \notag \\ &  -(X'_{qi}-X'_{di})I_{di}I_{qi} - D_{i}(\omega_{i}-\omega_{s}) \label{eq: dynamics_gen_d}  \\
    & \quad \quad \quad \quad \quad \quad \quad \quad \quad \quad \quad \quad \forall i= 1, \ldots, m, \nonumber
\end{align}
\ese
where $i$ is the generator index.
The state variables of the generator $i$ include
internal voltages in the $d$-axis and $q$-axis, $E'_{di}$ and $E'_{qi}$, the rotor angle, $\delta_i$, and  the angular velocity, $\omega_i$.
The parameters for each generator in (\ref{gen:dynamics_gen}) include transient open-circuit time constants of $d$-axis and $q$-axis, $T'_{do}$ and $T'_{qo}$, the machine inertia constant, $C$, the mechanical torque, $T_{M}$, the damping coefficient, $D$, the synchronous reference angular speed, $\omega_s$, and the synchronous and transient reactances in $d$-axis and $q$-axis, $X_{d}$, $X'_{d}$, $X_{q}$ and $X'_{q}$, respectively.
$E_{fdi}$ is the state variable of the exciter of the generator $i$, namely the scaled field voltage. 
Additionally, applying Kirchhoff's Voltage Law, we obtain the following stator algebraic equations:
\bse
\label{Kirshoff}
\beqn
    E'_{di}-V_{i}\sin(\delta_{i}-\theta_i)-R_{si}I_{di}+X'_{qi}I_{qi} = 0, \label{eq: Kirshoff_a}\\
    E'_{qi}-V_{i}\cos(\delta_{i}-\theta_i)-R_{si}I_{qi}-X'_{di}I_{di} = 0,
    \label{eq: Kirshoff_b}\\
    \forall  i=1,\ldots,m. \notag
\eeqn
\ese

\subsubsection{Dynamics of exciters, turbines, and speed governors model}
\label{exciter_model}
\begin{figure}[http]
    \centering
     \vspace{-\topsep}
    \includegraphics[width = 0.42\textwidth]{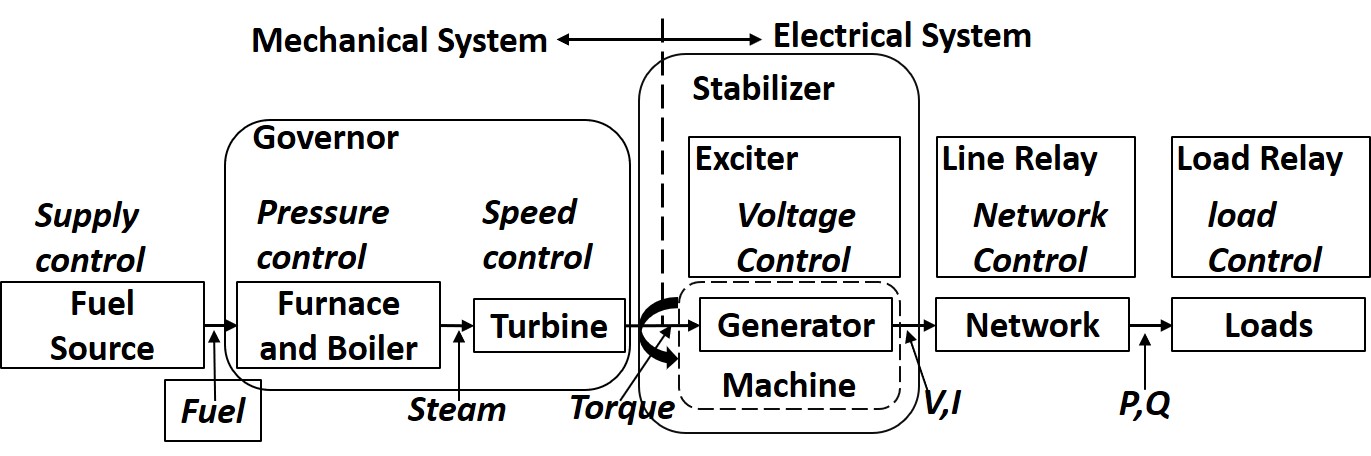}
    \vspace{-0.5em}
    \caption{Generator with exciter and governor \cite{PowerWorld} (page 2).}
    \label{fig:physical_structure}
\end{figure}

A generator has a voltage stabilizer/exciter for voltage control and a governor for speed control as depicted in Figure \ref{fig:physical_structure}.

\begin{figure}[b]
    \centering
    \includegraphics[width = 0.42\textwidth]{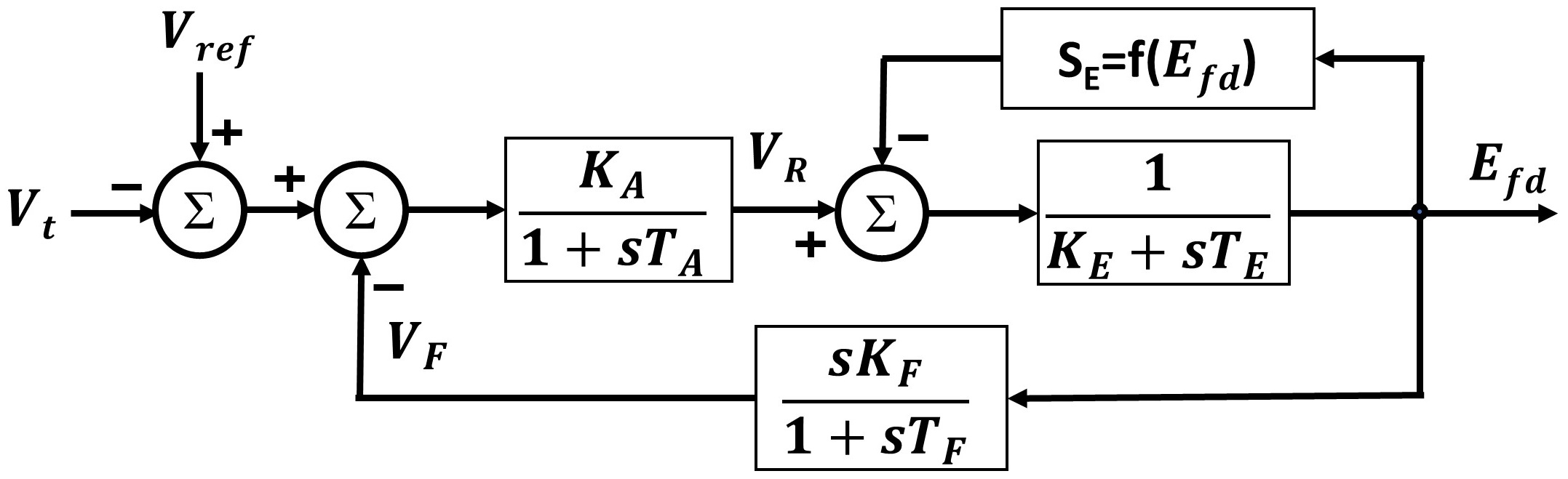}
        \vspace{-0.5em}
    \caption{IEEE type 1 Exciter system \cite{sauer2017power} (page 218).}
    \label{fig:exciter}
\end{figure}

The IEEE type 1 exciter regulates the generator's voltage, $E_{fdi}$ in (\ref{gen:dynamics_gen}),
and has the following dynamic model:
\bse
\label{exciter:dynamics}
\begin{align}
    &T_{Ei}\frac{\D E_{fdi}}{\D t}  = -(K_{Ei}+S_{Ei}(E_{fdi}))E_{fdi}+V_{Ri} \label{eq: exciter_a}  \\
    &T_{Fi}\frac{\D R_{fi}}{\D t} = - R_{fi}+\frac{K_{Fi}}{T_{Fi}} E_{fdi},  \label{eq: exciter_b}~  \text{with}~ R_{fi} \triangleq \frac{K_{Fi}}{T_{Fi}}E_{fdi} - V_{Fi}, \\
    &T_{Ai}\frac{\D V_{Ri}}{\D t}  = -V_{Ri}+K_{Ai}R_{fi} \notag \\& \quad \quad \quad \quad \quad -\frac{K_{Ai}K_{Fi}}{T_{Fi}}E_{fdi}+K_{Ai}(V_{refi}-V_{i})  \label{eq: exciter_c}\\
    &\quad \quad \quad\quad \quad \quad \quad \quad \quad \quad \quad \forall i= 1, \ldots, m. \nonumber
\end{align}
\ese 
whose state variables include the scaled field voltage, $E_{fdi}$, the rate feedback, $R_{fi}$, and the scaled input to the main exciter, $V_{Ri}$ \cite{sauer2017power}.
The overall block diagram of the exciter is shown in Figure \ref{fig:exciter}.
Note $R_{fi}$ can be calculated using the scaled field voltage $E_{fdi}$ and the scaled output of the stabilizing transformer $V_{Fi}$. 
The exciter's parameters include the exciter gain, $K_{Ei}$, the exciter time constant, $T_{Ei}$, exciter saturation, $S_{Ei}$, the rate feedback gain, $K_{Fi}$, the rate feedback time constant, $T_{Fi}$, the amplifier gain, $K_{Ai}$, and the amplifier time constant, $T_{Ai}$.
The exciter saturation $S_{Ei}$ is a function of $E_{fdi}$, which is set as $S_{Ei}(E_{fdi}) = 0.0039e^{1.555 E_{fdi}}$ as in \cite{sauer2017power}.

\begin{figure}[t!]
    \centering
    \includegraphics[width = 0.42\textwidth]{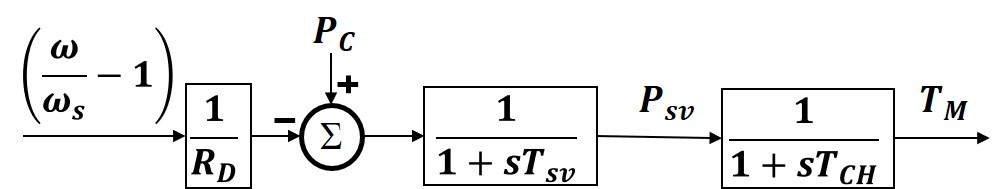}
    \caption{Turbine and speed governor model \cite{alvarado2001stability} (page 4).}
    \label{fig:governor}
\end{figure}

Figure \ref{fig:governor} shows the block diagram of the turbine and speed governor model, whose dynamics can be represented as follows:
\begin{subequations}
\label{governor:dynamics}
\begin{align}
    & T_{CHi} \frac{\D T_{Mi}}{\D t} = -T_{Mi} + P_{SVi} \label{eq: governor_a} \\
    & T_{SVi}\frac{\D P_{SVi}}{\D t}  = -P_{SVi} + P_{Ci} - \frac{1}{R_{Di}} \left(\frac{\omega_{i}}{\omega_{s}} - 1  \right) \label{eq: governor_b}  \\
    & \quad \quad \quad \quad \quad \quad \quad \quad \quad \quad \quad \quad \forall i= 1, \ldots, m. \nonumber
\end{align}
\end{subequations}
Parameters include the steam chest time constant, $T_{CHi}$, the steam valve time constant, $T_{SVi}$, and speed regulation quantity, $R_{Di}$.
Variable $P_{SVi}$ is the steam valve position. 
Variable $P_{Ci}$ is the power control continuously computed based on the given initial value set point $P^{0}_{Ci}$, a constant participation factor $k^{pf}_i$, and the total change in loads $Z$ $\big( \sum^m_i k^{pf}_i = 1 \big)$:
\beqn
    P_{Ci} = P^{0}_{Ci} + k^{pf}_i Z, \quad \qquad \forall i = 1, \ldots, m,
    \label{eq: power_control}
\eeqn

\subsubsection{Network power flow equations}
\label{network_balance}

For generator buses $(i = 1, \ldots,m)$, $I_{di}$, $I_{qi}$, since $V_i$, and $\theta_i$ are coupled with the state variables of internal generators' voltages, $E'_{di}$ and $E'_{qi}$, by KVL, the active and reactive power balances are:
\bse
\label{powerbalance}
\beqn
    I_{di}V_{i}\sin(\delta_{i}-\theta_{i})+I_{qi}V_{i}\cos(\delta_{i}-\theta_{i})  -P_{Li} \notag \\
    =\sum^{n}_{k=1}V_{i}V_{k}Y_{ik}\cos(\theta_{i}-\theta_{k}-\alpha_{ik})\label{eq: powerbalance_a}\\
    I_{di}V_{i}\cos(\delta_{i} -\theta_{i})-I_{qi}V_{i}\sin(\delta_{i}-\theta_{i}) - Q_{Li} \notag \\  =\sum^{n}_{k=1}V_{i}V_{k}Y_{ik}\sin(\theta_{i}-\theta_{k}-\alpha_{ik}) \label{eq: powerbalance_b} \\
    \quad \quad \quad \quad i = 1,\ldots,m, \nonumber
\eeqn
where  $I_{di}V_{i}\sin(\delta_{i}-\theta_{i})+I_{qi}V_{i}\cos(\delta_{i}-\theta_{i})$ is the active power and  $I_{di}V_{i}\cos(\delta_{i} -\theta_{i})-I_{qi}V_{i}\sin(\delta_{i}-\theta_{i})$ is the reactive power, generated by the generator.
Also, $P_{Li}$ and $Q_{Li}$ are the active and reactive power demands respectively.
Similarly, for $n-m$ remaining load buses, the active and reactive power balances are:
\beqn
    P_{Li} =  -\sum^{n}_{k=1} V_{i}V_{k}Y_{ik}\cos(\theta_{i}-\theta_{k}-\alpha_{ik}) \label{eq: powerbalance_c}\\
    Q_{Li}  = -\sum^{n}_{k=1} V_{i}V_{k}Y_{ik}\sin(\theta_{i}-\theta_{k}-\alpha_{ik}) \label{eq: powerbalance_d} \\
     i = m+1,\ldots,n. \nonumber
\eeqn
\ese

\subsubsection{DAE compact form}
\label{DAE_compact}
 equations  \eqref{gen:dynamics_gen}-\eqref{powerbalance} together form the DAEs modeling power system dynamics in the compact form \cite{sauer2017power} as:
\begin{subnumcases}{\label{compact:form}}
    \dfrac{\D x}{\D t} = f(x,y) \label{dynamics}\\
    0 = g(x,y), \label{algebra} 
\end{subnumcases}
\vspace{-\topsep}
\begin{align}
& x = [x^\top_1, x^\top_2,\ldots,x^\top_m]^\top, \nonumber\\
& y = [I_{di}, I_{qi}, V_i, \theta_i ~~  (i =1,\ldots,n)]^\top, \nonumber
\end{align}
where \eqref{dynamics} represents the dynamics of $m$ generator and (\ref{algebra}) represents the KVL and power flow balances in the network.
The nonlinear vector-valued functions, $f$ and $g$, are derived from the state vector $x$, where $x_i =[E'_{di}, E'_{qi},\delta_i, \omega_i, E_{fdi}, R_{fi}, V_{Ri}, T_{Mi},P_{SVi}]^\top$ is the state vector for generator $i$ and vector of algebraic variables $y$.

\begin{figure}[b]
    \centering
    \includegraphics[width = 0.4\textwidth]{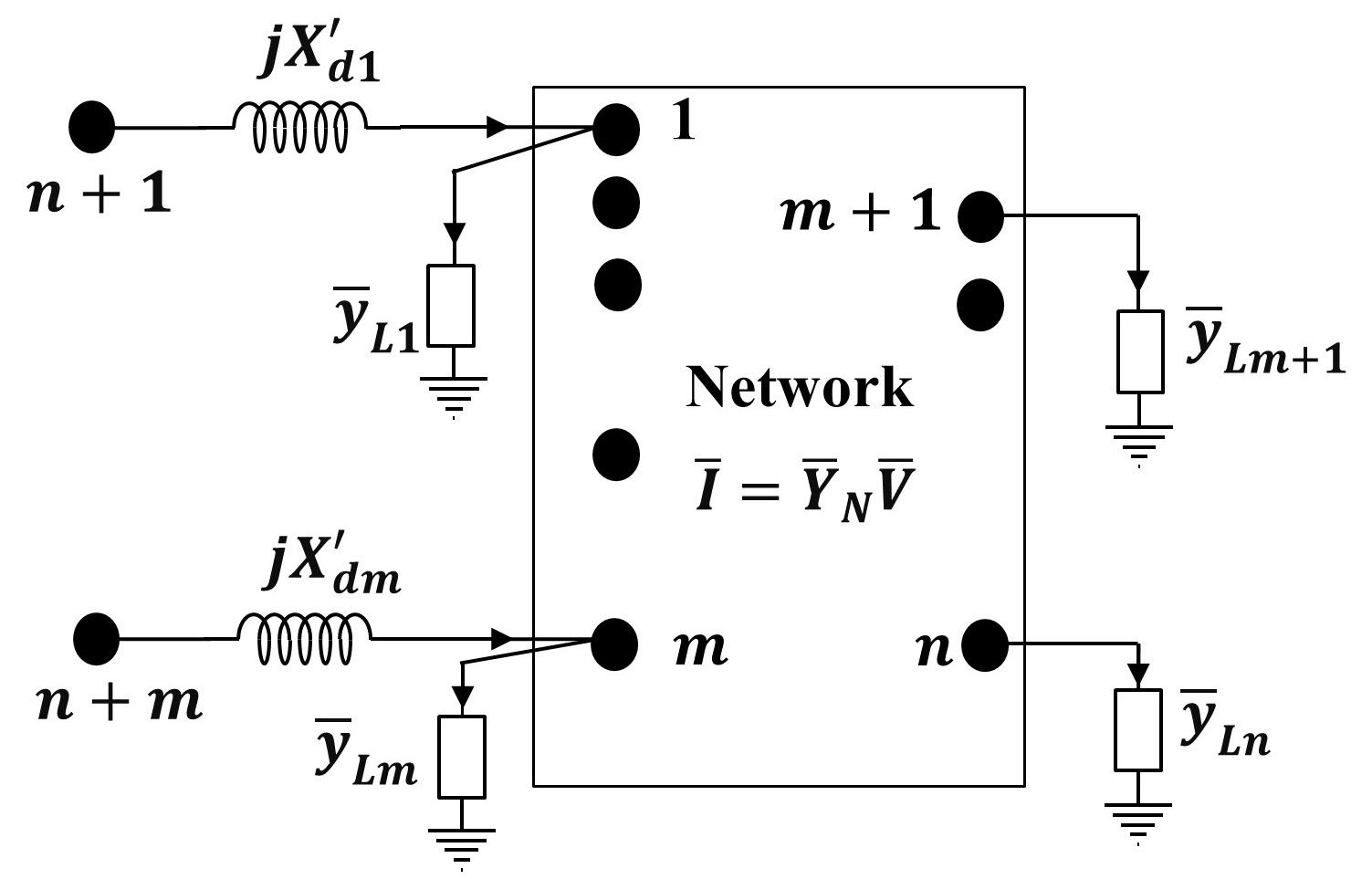}
    \caption{Internal-Node model \cite{sauer2017power} (page 178).}
    \label{fig:internal_model}
\end{figure}

\subsection{Special Case: Internal-Node model}

A special case in which the DAEs of power system dynamics become ODEs is the internal-node model widely used in first-swing transient stability analysis \cite{sauer2017power}.
It is also called the network with constant impedance as the nodal demand is represented by an admittance as:
\begin{equation}
    \overline{y}_{Li} = \frac{-(P_{Li}-jQ_{Li})}{V^2_i}, \quad \forall i = 1,\ldots,n.
\end{equation}
Assuming the damper-winding constants are very small and $T'_{qoi}$ and $D_i$ are zero, so $E'_{di}$ can be substituted to \eqref{eq: dynamics_gen_d} and \eqref{eq: dynamics_gen_a} is eliminated.
Additionally, assuming $T'_{doi} = \infty$ and $X_{qi} = X'_{di}$, $E'_{qi}$ becomes a constant equal to the initial value $E_{i}$.
Thus, \eqref{eq: dynamics_gen_b} is eliminated and the $n$ bus transmission network is augmented at the generator buses $1,\ldots,m$.
These generator internal buses are denoted as $n+1,\ldots,n+m$.
Consequently, we can represent the dynamic network of the power grid as in Figure \ref{fig:internal_model} where nodal demand is represented by an admittance $\overline{y}_{Li}$.
The network equation becomes:
\beqn
    \begin{bNiceMatrix}[first-col]
        {m} & { \bar{I}_A} \\
        {n} & {0}
    \end{bNiceMatrix} 
    = \begin{bNiceArray}{cc}[first-row,first-col]
        & m &  n \\
        m & {\overline{Y}_A} &  {\overline{Y}_B} \\
        n &  {\overline{Y}_C} &  {\overline{Y}_D}
    \end{bNiceArray} \begin{bNiceMatrix}
         { \overline{E}_A} \\
        {\overline{V}_B}
    \end{bNiceMatrix},
\eeqn
where $\overline{Y}_A = \overline{y}$, $\overline{Y}_B = \Big[ -\overline{y} ~|~ 0  \Big]$, $\overline{Y}_C = \Big[ \dfrac{-\overline{y}}{0} \Big]$, $\overline{Y}_D = \overline{Y}_N + \begin{bmatrix}
\overline{y} & 0 \\ 0 & 0 \end{bmatrix} + \begin{bmatrix}
Diag(\overline{y}_{Li}) & 0 \\
0 & 0
\end{bmatrix}$, and $\overline{y}=Diag \left( \frac{1}{jX'_{di}}\right)$.

Since there is no current injection, the $n$-network buses can be eliminated, and we have:
\begin{align}
    \overline{I}_A = (\overline{Y}_A - \overline{Y}_B\overline{Y}_D^{-1}\overline{Y}_C)\overline{E}_A  = \overline{Y}_{int}\overline{E}_A,
\end{align}
where the elements of $\overline{I}_A$, $\overline{E}_A$, and $\overline{Y}_{init}$ are $\overline{I}_i = \big( I_{di}+jI_{qi}e^{j(\delta_i-\pi/2)} \big)$, $\overline{E}_i = E_i \angle \delta_i$, ${\overline{Y}_{ij}} = {G_{ij}} {+} {j}{B_{ij}}$.
The network now has only $m$ internal nodes, and the KVL for the network are:
\begin{equation}
    \overline{I}_i = \sum^m_{j=1} \overline{Y}_{ij}\overline{E}_j, \quad \forall i = 1,\ldots,m,
\end{equation}
By denoting $\delta_i - \delta_j \triangleq \delta_{ij}$, the real electrical power output at $i^{th}$ internal node in Figure \ref{fig:internal_model} is given by
\begin{align}
    P_{ei} & = Re \left[ \overline{E}_i\overline{I}^*_i \right] = Re \left[ E_ie^{j\delta_i}\sum^m_{j=1}\overline{Y}^*_{ij}\overline{E}^*_j \right] \notag \\
    & = E^2_iG_{ii} + \sum^m_{j=1, j\neq i} (C_{ij}sin(\delta_{ij})+D_{ij}cos(\delta_{ij})) \label{eq: poweroutput}\\
    &~~~~~~\text{with} ~~C_{ij} = E_i E_j B_{ij}, D_{ij} = E_i E_j G_{ij}, \nonumber ~ \forall i = 1, \ldots, m. \nonumber
\end{align}

For the internal node model, we do not have the state variables for voltages and currents.
Therefore, the exciter systems are not used, and the model includes the generators and governor systems.
Consequently, the power system dynamics in the Internal-Node model become simple ODEs as follows \cite{sauer2017power}:
\bse
\label{internal_model}
\begin{align}
    \frac{\D \delta_{i}}{\D t} &= \omega_{i} - \omega_{s}  \label{eq: internal_model_a}\\
    \frac{2H_{i}}{\omega_{s}} \frac{\D \omega_{i}}{\D t} &= T_{Mi} -  P_{ei} \label{eq: internal_model_b} \\
    T_{CHi} \frac{\D T_{Mi}}{\D t} & = -T_{Mi} + P_{SVi} \label{eq: internal_model_c} \\
    T_{SVi}\frac{\D P_{SVi}}{\D t}  = -P_{SVi} + & P_{Ci} - \frac{1}{R_{Di}} \left(\frac{\omega_{i}}{\omega_{s}} - 1  \right) \label{eq: internal_model_d} \\
    &  \quad \quad \quad \quad \forall i= 1, \ldots, m, \nonumber
\end{align}
\ese
where we recall that $P_{Ci} = P^0_{Ci} + k^{pf}_i Z$ and $P_{ei}$ is given in (\ref{eq: poweroutput}).

\subsection{Converting DAEs system into ODEs system}

Standard numerical integration techniques of ODEs are generally more efficient and accurate than ones developed for DAEs \cite{hairer1993solving}.
Therefore, it is required to transform a DAE system into an equivalent set of ODEs:
\beqn
\dfrac{\D z}{\D t} = \hat{f} (t,z), \label{standardODEs}
\eeqn
where $z$ is a new set of state variables so that it can be solved \cite{ascher1998computer}. 
Unfortunately, transforming DAEs of power system dynamics into equivalent ODEs is a challenging task due to the complexity of nonlinear power flow equations embedded in (\ref{algebra}).
One standard transformation method is modeling algebraic variables as explicit functions of state variables.
Specifically, if the Jacobian $\dfrac{\partial g}{\partial y}$ is nonsingular, there exists a function $\hat{g}$ such that \eqref{algebra} can be equivalently cast as $y = \hat{g}( x)$ in accordance with the implicit function theorem \cite{krantz2002implicit}.
In general, the nonsingularity of $\dfrac{\partial g}{\partial y}$ holds for normal operations since the power flow and Kirchhoff's voltage equations have a unique solution.  
Thus, substituting $y$ by $\hat{g}(x)$, the formulation (\ref{compact:form}) can be rewritten in the standard form (\ref{standardODEs}), i.e.,$
    \dfrac{\D x}{\D t} = f( x,\hat{g}(x)) = \hat{f}(x)$.
However, it is very difficult to explicitly construct the magnitudes and phases of nodal voltages as a function $\hat{g}$ of state variables of generators, particularly injected currents and internal voltages.
Moreover, in some operating scenarios, the Jacobian $\dfrac{\partial g}{\partial y}$ can be singular \cite{grijalva2005necessary} and $\hat{g}(x)$ does not exist.

The difficulty of solving a DAE system depends on its index. 
DAEs of index 0 or 1 can be easily converted into equivalent ODEs.
Nevertheless, given a large number of algebra equations and variables in (\ref{algebra}), the power system dynamics' DAEs (\ref{compact:form}) can have an index greater than two, which is very difficult to solve   \cite{petzold1982differential}.
To overcome these issues, we employ the Pantelides algorithm \cite{Panthe}, which systematically reduces high-indexed DAEs to lower-indexed ones by selectively adding differentiated forms of equations that are already present in the system \cite{Panthe}. 
Consequently, we can convert the power system dynamics' DAEs (\ref{compact:form}) into an indexed-1 DAE system and then equivalent ODEs (\ref{standardODEs}), where $z$ encapsulates original state variables $x$ and parts of algebraic variables $y$.

Once we obtain the ODE formulation (\ref{standardODEs}), we can employ numerical discretization methods such as the Forward Euler method to solve it on classical computers \cite{ascher1998computer}:
\begin{align}
 \dfrac{{z(t+\Delta)} {-} {z(t)}}{\Delta } \approx  \hat{f}(z(t))
 \iff  {z(t+\Delta)} {=} {z(t)} {+} {\Delta \hat{f}(z(t)).}  \label{euler}
\end{align}

\section{QUANTUM COMPUTING METHOD FOR POWER SYSTEM DYNAMICS' DAEs}
\label{Quantum}
\subsection{Overview of quantum computing}
\subsubsection{Qubits and vector representation} 
\begin{figure}[t!]
    \centering
    \includegraphics[width = 0.4\textwidth]{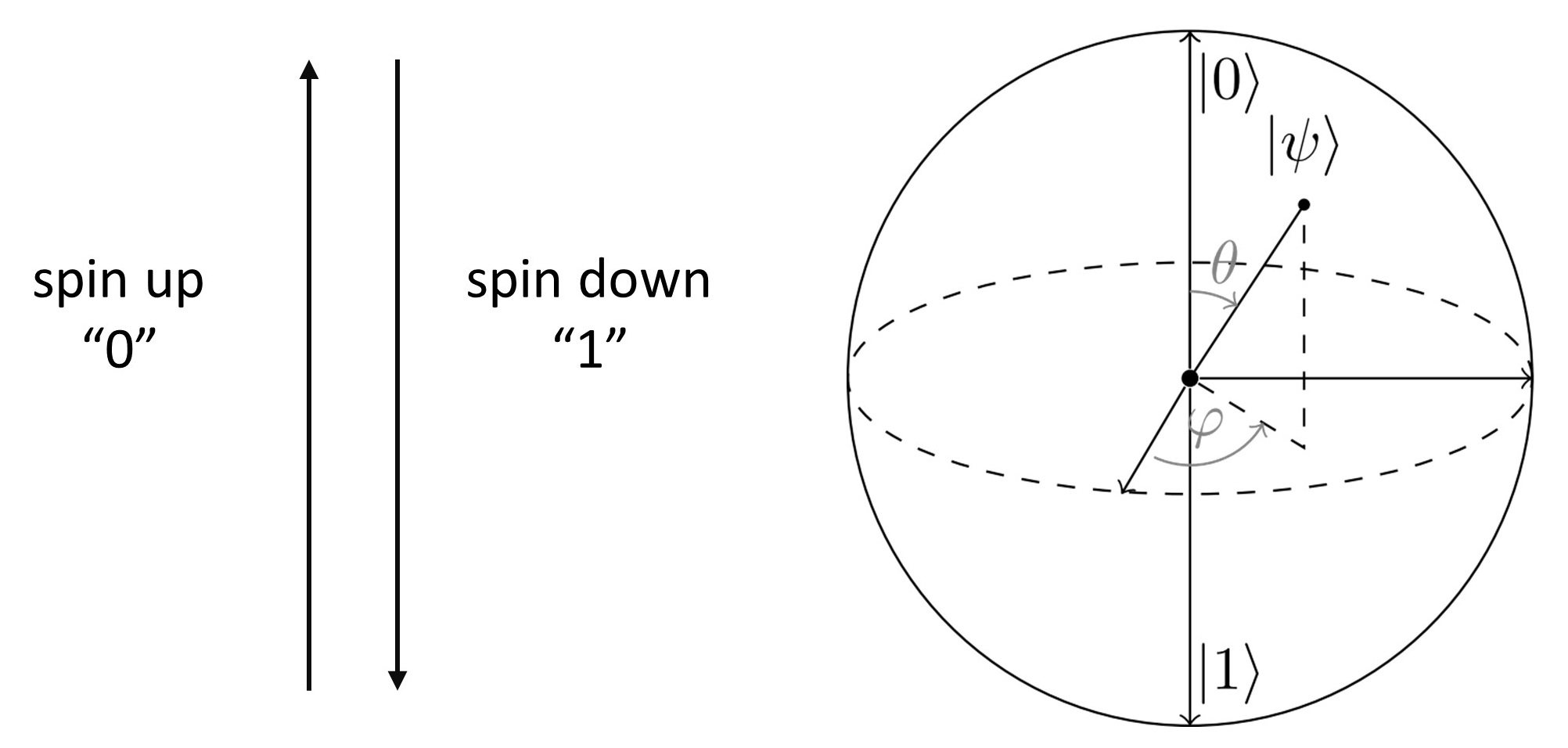}
    \caption{The particle spin can represent 0
and 1. The Block Sphere with $\ket{0}$ and $\ket{1}$ at the poles represents a qubit where $\alpha = \cos\frac{\theta}{2}$, $\beta = e^{i \phi} \sin\frac{\theta}{2}$ \cite{lapierre2021introduction} (page 67).} 
    \label{fig:qubit}
\end{figure}

Quantum computers represent information (digital bits, 0 and 1) using two different states of a quantum system \cite{lapierre2021introduction}, e.g., two different quantized energy levels (0 as the ground state and 1 as an excited state of an electron) or two spin states of the particle (0 as electron spins ``down'' and 1 as electron spins ``up''). 
Thus, a qubit depicted in Figure \ref{fig:qubit} can be considered as the supposition of two quantum states. 
Using Dirac ket and bra notation of vector, i.e., $\ket{}$ and $\bra{}$, (see Appendix \ref{ket_bra}), it can be mathematically modeled as follows: 
\beqn
\ket{\psi} = \alpha \ket{0} + \beta \ket{1} \longrightarrow \psi =  \begin{bmatrix}
          \alpha \\
           \beta
         \end{bmatrix} \in \mathbb{C}^2,
\eeqn
where complex values $\alpha$, $\beta$ are probability amplitudes satisfying the
normalization condition $ |\alpha|^2 + |\beta|^2 =1 $, i.e., 
we can obtain 0 with probability $ |\alpha|^2 $ and 1 with probability $|\beta|^2$ if measuring the quantum state $\ket{\psi}$. 
Thus, a single qubit can represent a column vector $\psi = \left[ \alpha, \beta \right]^\top$ in a 2-dimensional complex vector space with two computational basis states:
\beqn
\ket{0} \longrightarrow \begin{bmatrix}
          1 \\
           0
         \end{bmatrix},~~~~
\ket{1} \longrightarrow \begin{bmatrix}
          0 \\
           1
         \end{bmatrix}.
\nonumber
\eeqn

\noindent
Two qubits can represent a vector in 4-dimensional complex vector spaces using the following 4 computational basis states:
\beqn
\ket{00} \rightarrow \begin{bmatrix}
          1 \\
           0\\
           0\\
           0
         \end{bmatrix},~
\ket{01} \rightarrow \begin{bmatrix}
          0 \\
           1\\
           0\\
           0
         \end{bmatrix},
\ket{10} \rightarrow \begin{bmatrix}
          0 \\
           0\\
           1\\
           0
         \end{bmatrix},
\ket{11} \rightarrow \begin{bmatrix}
          0 \\
           0\\
           0\\
           1
         \end{bmatrix},
\nonumber
\eeqn
which can be re-denoted as $\ket{0}, \ket{1}, \ket{2}, \ket{3}$ for convenience.

In general, a set of $n$ qubits can represent a column vector of $2^n$ complex elements using $2^n$ Hilbert space computational basis states $\ket{j},~ j=0, \ldots {2^n}{-}{1}$. 
In other words, a column vector of $N$ element requires at least $n= \log_2N$ qubits for representation. 
Note the superposition of these $2^n$ states, i.e., these states can be input simultaneously, is known as quantum parallelism, enabling quantum algorithms to outperform classical ones for certain problems, including solving/updating linear equations.  
These remarks can be utilized for linear updating state variables and polynomially approximating nonlinearity in power system dynamics.

\subsubsection{Time evolution of a quantum system} \label{time_evolution}
 the time evolution of quantum state $\ket{\psi}$ at time $t$ under the following time-independent Schrödinger equation \( i \frac{\partial}{\partial t} \ket{\psi(t)} = H \ket{\psi(t)} \) is $\ket{\psi(t)} = e^{-i H t} \ket{\psi(0)}$ where $H$ is the Hamiltonian operator (a Hermitian matrix acting on $n$ qubits) \cite{lapierre2021introduction}.
The exponential component $e^{-i H t}$ can be expressed using a Taylor expansion as follows:
\beqn
e^{-iHt} = \sum\limits_{j=0}^\infty \dfrac{(-i H t)^j}{j!},
\eeqn
thus, the time evolution of the quantum state $\ket{\psi}$ of $n$-qubits from the initial state $\ket{\psi(0)}$ can be represented as:
\beqn
\ket{\psi(t)} = e^{-i H t} \ket{\psi(0)} = \sum\limits_{j = 0}^\infty \dfrac{(-i H t)^j}{j!} \ket{\psi(0)}, \label{HalmintonOriginal}
\eeqn
which will be utilized for modeling the evolution of nonlinear state functions in power system dynamics.

\subsection{Application on solving ODEs in power system dynamics} 
We aim to leverage the potential application of advances in quantum computing for solving linear equations \cite{harrow2009quantum} for tackling the power system dynamics of a large-scale network.
The Forward Euler equation (\ref{euler}) is indeed a linear equation that can be rewritten as:
\beqn
    \ket{z(t+\Delta)} = \ket{z(t)} + \Delta \ket{\hat{f}(z(t)}. \label{eq:principle}
\eeqn
To implement (\ref{eq:principle}) in a quantum computer, we need to construct the quantum state $\ket{z(t)}$ and $\ket{\hat{f}(z(t)}$.
To this end, we employ Leyton-Osborne's algorithm, which includes the following steps \cite{leyton2008quantum}.

\subsubsection{Data encoding of state vector}
We employ the amplitude encoding to encode the state vector $z$ of length $N$ in equation (\ref{euler}) into amplitudes of an $n$-qubit quantum state, where $n = \log_2N+1$. 
First, $z$ must be normalized to ensure that $\sum_{j=1}^N |z_j|^2 =1$ since its elements now represent the amplitudes of a quantum state. 
Then, we map the variables $z_j(t)$ to the quantum state as follows:
\beqn
    \ket{{z}} = \frac{1}{\sqrt{2}}\ket{0} + \frac{1}{\sqrt{2}} \sum_{j=1}^{N} {z}_j\ket{j}, \label{eq:encode}
\eeqn
where $\ket{j}$ is the Hilbert space computational basis.

Now, the task is to create the quantum state $\ket{\hat{f}(z(t))}$ comprising of $\log_2N + 1$ qubits so equation (\ref{eq:principle}) can be conducted in a quantum computer.
This can be accomplished in quantum computers using a combination of Taylor expansion, mathematical tensor, and Hamiltonian simulation, which will be elaborated on next.  

\subsubsection{Polynomial approximation of state function}
The $j^{th}$ element in $\hat{f}(z(t))$ can be polynomially approximated by the second-order Taylor expansion as follows:
\begin{align}
\hat{f}_{j}(z(t)) \approx  &\hat{f}_{j}(\bar{z}) + \frac{\partial \hat{f}_j(z(t))}{\partial z} ( z(t) - \bar{z}) \nonumber \\
& + \frac{1}{2}( z(t) - \bar{z})^\top \nabla^{2}\hat{f}_j(z(t))( z(t) - \bar{z}),
\end{align}
which can be compactly rewritten in the form:
\begin{gather}
\hat{f}_{j}(z(t)) = \underbrace{\hat{f}_{j}(\bar{z})}_{a^j_{0,0}} + \sum\limits_{k=1}^{N} a^j_{0,k} z_k(t)
+ \sum\limits_{v,k=1}^N  a^j_{v,k} z_v(t) z_k(t)
\label{taylor_expand}
\nonumber
\end{gather}
where $z_k(t), z_v(t)$ are the $v^{th}$ and $k^{th}$ elements of the vector $z(t) - \bar{z}$, $a^j_{v,k}, (v=0,\ldots, N, k= 0,\ldots N)$ are coefficients of the polynomial approximation of $\hat{f}_{j}(z(t))$.
Here, $\bar{z}$ represents approximating points (operating or initial state of the power systems) where the state function values at $\bar{z}$, i.e., $\hat{f}_{j}(\bar{z})$ is already determined and can be represented as $a^j_{0,0}$.
After adding extra variable $z_0 = 1$, we obtain the following polynomial form of $\hat{f}_j$: 
\beqn
\hat{f}_{j}(z(t)) = \sum\limits_{v,k=0}^N a^j_{v,k} z_v(t) z_k(t) \label{taylor_compact}
\eeqn

\subsubsection{Nonlinear transform for the state function} 
According to equation (\ref{taylor_compact}), computing $\hat{f}_{j}(z(t))$ requires encoding all monomials $z_v(t) z_k(t),\forall v,k$.
In complex linear algebra, these values are given in the tensor product as illustrated as follows:

\includegraphics[width = 0.46\textwidth]{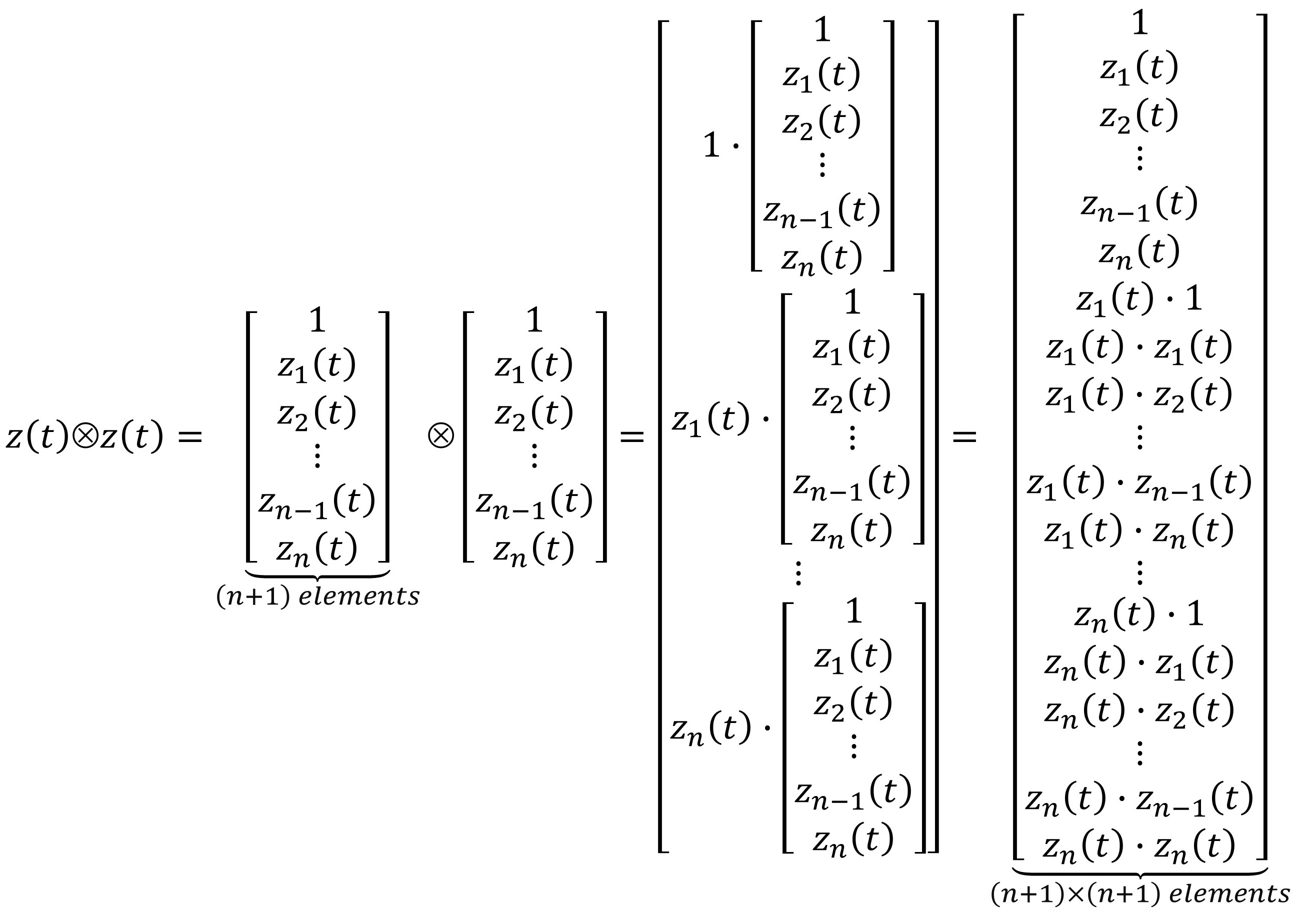}

\noindent
This means we can capture all values $z_v(t) z_k(t),\forall v,k$ in the probability amplitudes of the tensor product of two copied quantum states if making a copy of quantum state is possible:
\beqn
    \ket{z} \ket{z} = \frac{1}{2}\sum^{N}_{v,k=0} z_v z_k \ket{vk}. \label{tensor}
\eeqn
To assign corresponding coefficients, $a^j_{v,k}$, to each monomial $z_v z_k$, we define the operator $A$ as follows:
\beqn
A = \sum\limits_{j,v,k=0}^N a^j_{v,k} \ket{j 0} \bra{vk}. 
\eeqn
Acting on $\ket{z} \ket{z}$, we obtains the information of $\hat{f}(z)$:
\beqn
A \ket{z} \ket{z} = \frac{1}{2} \sum\limits_{j,v,k=0}^N a^j_{v,k} z_v z_k \ket{j} \ket{0} = \frac{1}{\sqrt{2}} \ket{\hat{f}(z)} \ket{0}. \label{test}
\eeqn 
The formulation (\ref{test}) shows that we can encode $\ket{\hat{f}(z)}$ using quantum state $\ket{z}$ by simulating $A \ket{z} \ket{z}$ in a quantum computer.
Unfortunately, making a copy of a quantum state is physically difficult due to the non-cloning theory.
This can be overcome using the Hamiltonian simulation and exploiting the truncated Taylor series of the matrix exponential, which will be discussed next.

\subsection{Simulating state function in a quantum computer}
We employ the well-known Von-Neumann measurement prescription by constructing the following Hamiltonian \cite{peres1997quantum}:
\beqn
H = i A \otimes \ket{1}_P \bra{0} - i A^\dag \otimes \ket{0}_P \bra{1} \label{halminton}
\eeqn
where $\otimes$ denote a tensor product,  $A^\dag$ is the adjoin of $A$, and $P$ is the qubit pointer  \cite{lawrence2023pointers,zurek1981pointer}. 
If we simulate the Hamiltonian $H$ in (\ref{halminton}) with the initial state $\ket{z} \ket{z} \ket{0}_P$ in a quantum computer, the quantum system will evolve in accordance with $H$ for a time step $\epsilon$ following (\ref{HalmintonOriginal}) to reach the following steady-state $\ket{\Psi}$ \cite{berry2007efficient} (see Appendix \ref{Steady-state}):
\begin{align}
\ket{\Psi} &= e^{-i \epsilon H} \ket{z} \ket{z} \ket{0} = \sum\limits_{j=0}^\infty \frac{ (-i \epsilon H)^j}{j!} \ket{z} \ket{z} \ket{0} \nonumber\\
             &= \ket{z} \ket{z} \ket{0} + \epsilon A \ket{z} \ket{z} \ket{1} - \ldots \label{steady_state}
\end{align}
where the second term in its truncated Taylor expansion contains $A \ket{z} \ket{z}$.
Hence, to produce $\ket{\hat{f}(z)}$, we need to measure the state $\Psi$ of the $n$-qubits and post-select on the computation basis $\ket{1}$.

Although both $\ket{z(t)}$, and $\ket{\hat{f}(z(t0)}$ are encoded in quantum computers, we cannot operate ``+'' directly on quantum states, e.g., it is physically impossible to merge the probability amplitudes of spin states of two electrons. 
To perform the Forward Euler linear update \eqref{euler} in a quantum computer, we have to reformulate it as a set of linear equations in the form $M \ket{s} = \ket{b}$, i.e., 
\begin{equation}
    \underbrace{\begin{bmatrix}
    \mathbbm{1} & 0 \\
    -\mathbbm{1} & \mathbbm{1}
\end{bmatrix}}_{\text{\normalsize $M$}} \underbrace{\begin{bmatrix}
    s_0 \\ s_1 
\end{bmatrix}}_{\text{\normalsize $\ket{s}$}} = \underbrace{\begin{bmatrix}
    z(t) \\ \Delta \hat{f}(z(t))
\end{bmatrix}}_{\text{\normalsize $\ket{b}$}}, \label{eq:linearequivalent}
\end{equation}
 and employing quantum linear equation solvers, e.g., the HHL algorithm \cite{harrow2009quantum}.
 Obviously, the solution of equivalent linear equations (\ref{eq:linearequivalent}) results in $s_0 = z(t)$ and $s_1 =z(t) + \Delta \hat{f}(z(t)) = z(t+\Delta)$.

Thus, after preparing the quantum states $\ket{z(t)}$ and $\Delta \ket{\hat{f}(z(t))}$, we simply 
set up quantum states $\ket{b}$ as the concatenation of $\ket{z(t)}$ and $\ket{\hat{f}(z(t))}$ and perform HHL with matrix $M$ and $\ket{b}$.
The values of $\ket{z(t+\Delta)}$ can be extracted from solution $\ket{s}$ as the outcomes of the HHL algorithm (see Appendix \ref{HHL_algorithm}).

\begin{figure}[t!]
    \centering
    \includegraphics[width = 0.35\textwidth]{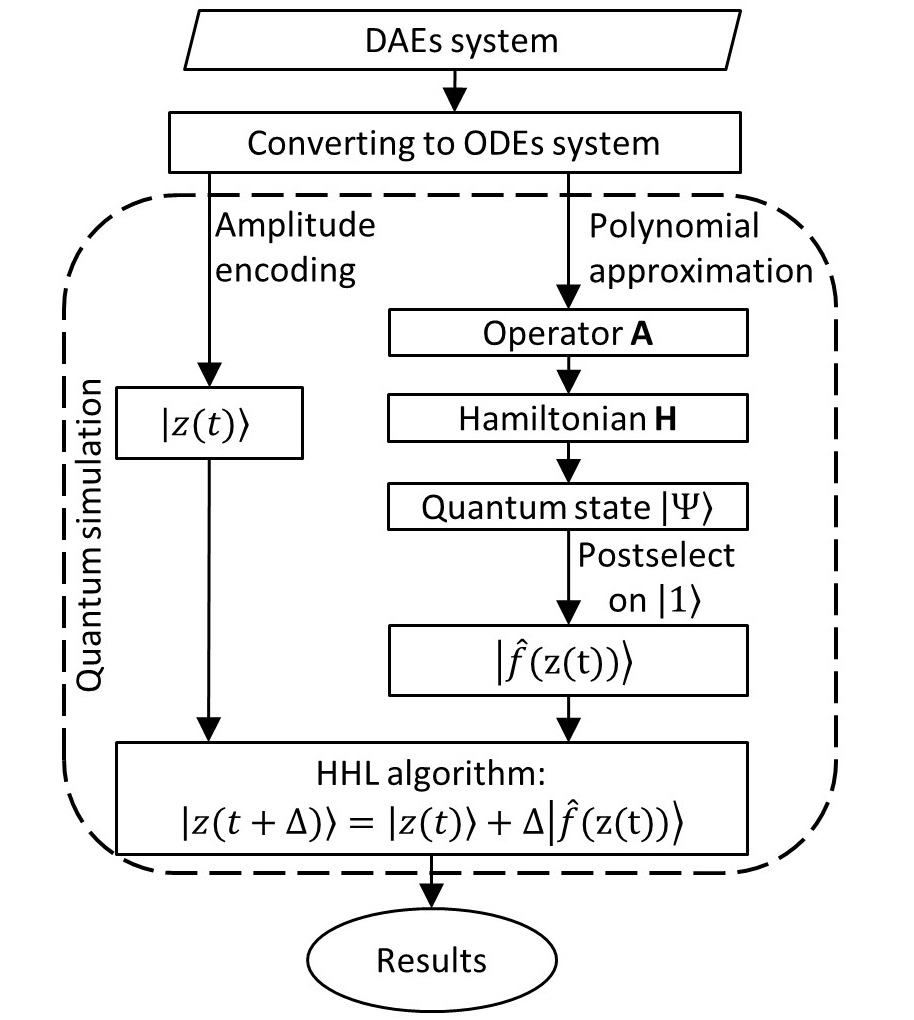}
    \caption{Flowchart for solving DAEs system by quantum computing.}
    \label{fig:Flowchart}
\end{figure}

In summary, the flowchart for solving the DAEs system by quantum computing is shown in Figure \ref{fig:Flowchart} consisting of three steps: encoding state variables $\ket{z(t)}$ via amplitude encoding, encoding nonlinear function $\ket{f(z(t))}$ using the tensor product of quantum states and Hamilton simulation, and finally linear update of state variable using HHL algorithm.

\section{IMPLEMENTATION AND RESULTS}
\label{implement_and_results}
\begin{figure}[t!]
    \centering
    \includegraphics[width = 0.47\textwidth]{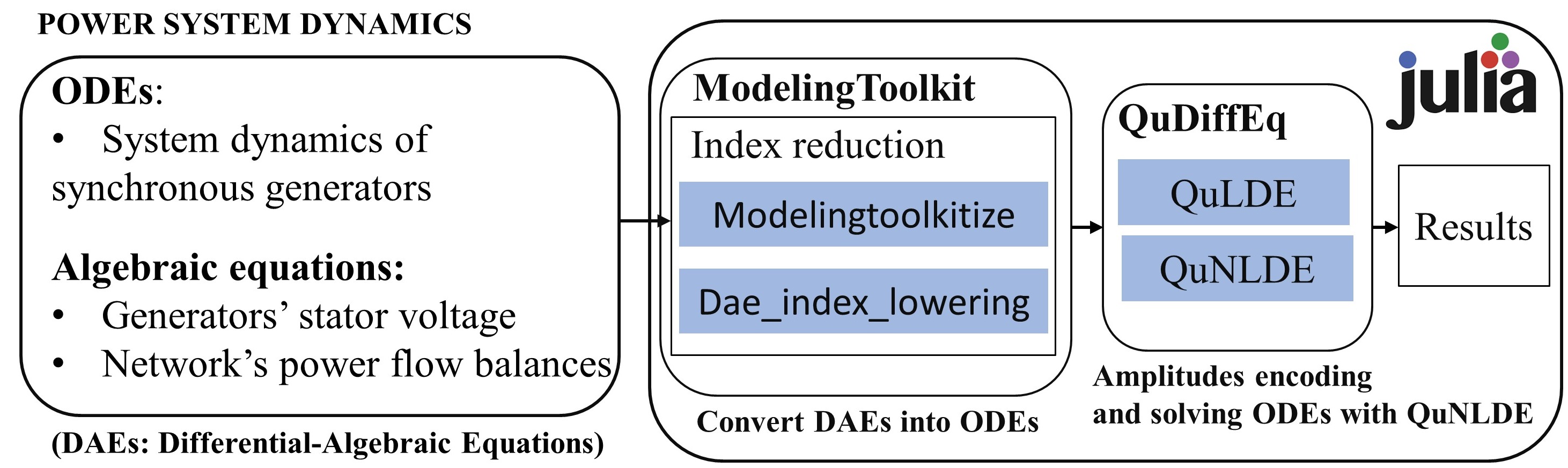}
    \caption{Implementation workflow (The QuNLDE solver employs the Leyton-Osborne algorithm presented in Section \ref{Quantum} \cite{leyton2008quantum}).}
    \label{fig:workflow}

\end{figure}

We utilize recent advances in quantum simulation and scientific machine learning tool-kits embedded in the Julia Computing environment for implementation, as shown in Figure \ref{fig:workflow}.
In particular, the DAEs that model power system dynamics are implemented in Julia via the ModelingToolkit (MTK) package \cite{ma2021modelingtoolkit}.
The set of DAEs in the form of symbolic representation is then converted into the index-1 DAE form and then the standard ODE form, i.e., formulation \eqref{compact:form}, by using the Pantelides algorithm with \textit{dae\_index\_lowering} function. 
The obtained ODEs can be solved by numerical methods (e.g., the Forward Euler algorithm) in classical computers.

For quantum computing, we solve the ODEs obtained by MTK with the QuDiffEq package, which supports Taylor expansion, amplitude encoding, and leveraging Yao.jl \cite{YaoFramework2019} as a quantum simulator.
Yao.jl is an efficient open-source framework for quantum algorithm design that can feature generic and differentiable programming for quantum circuits and support the simulation of small to intermediate-sized quantum circuits with state-of-the-art performance.  
Yao.jl offers quantum block intermediate representation, which is a tensor representation of quantum operations, i.e., quantum circuits and quantum operators (quantum gates, Hamiltonian, or the whole program).
Then, we employ Leyton-Osborne's algorithm embedded in the QuNLDE solver of QuDiffEq package \cite{leyton2008quantum} to solve examples of power system dynamic models given in \cite{sauer2017power}.

The dynamics of the power system are solved by quantum computing through two case studies; (i) a single-machine infinite bus system and (ii) a Western System Coordinating Council (WSCC) three-machine nine-bus system \cite{sauer2017power}.
The time step for linearization is set to 0.01s.  
The root means square error method is used to measure the difference between the quantum and classical methods.
A solid line represents the values of quantum computing, and a dashed line represents the values of the classical method.
The complete implementation details and source code of this study are available on the GitHub repository \cite{PowerSyatemDynamicsEthan}.

\subsection{Single-machine infinite bus system}
\label{single}

\begin{figure}[t!]
    \centering
    \includegraphics[width = 0.7\linewidth]{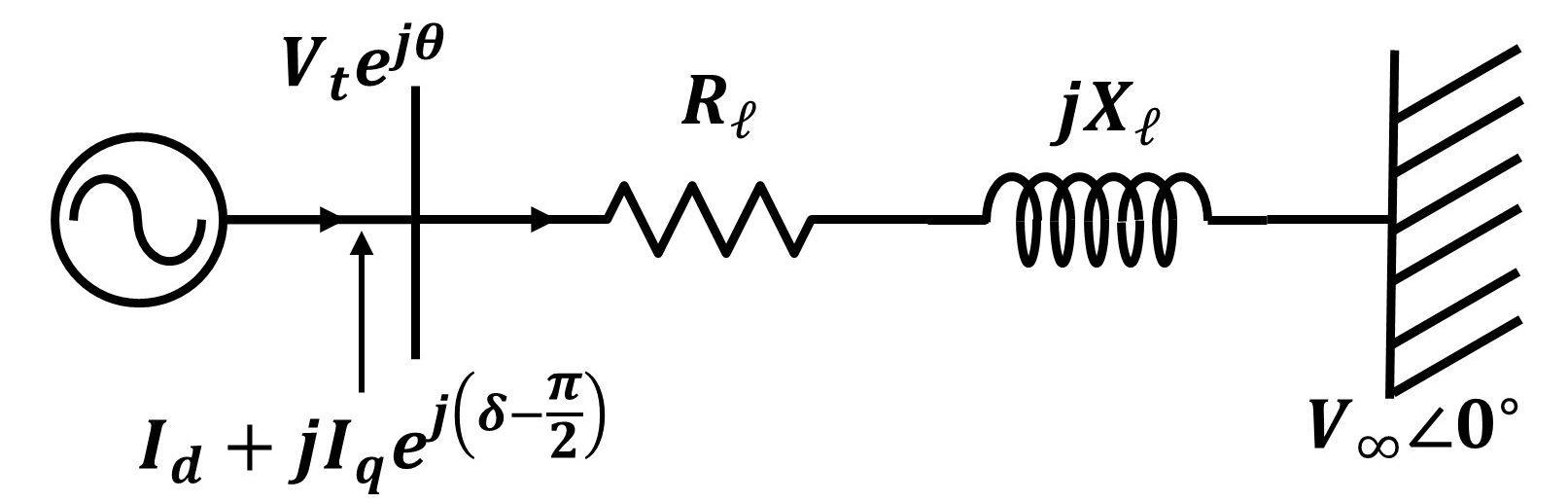}
    \caption{Single-machine infinite bus system \cite{sauer2017power} (page 210).}
    \label{fig:single}
\end{figure}

The Single-Machine Infinite Bus (SMIB) system, as illustrated in Figure \ref{fig:single}, presents a simplified representation of the power grid that is commonly used to study the dynamic behavior and transient stability of the power system.
Its dynamics can be modeled in the simple ODE form \cite{sauer2017power} (Chapter 5.8, pages 90-93):
\begin{subnumcases}{\label{SMIB:ODEs}}
    \frac{d\delta}{dt} = \omega - \omega_s \\
    \frac{d(\omega - \omega_s)}{dt} = K_1 - K_2\sin(\delta) - K_3(\omega - \omega_s),
\end{subnumcases}
where $\delta$ is the machine's internal voltage angle, $\omega$ is the angular speed of the synchronous machine at time $t$, $\omega_s$ is the reference angular speed, and $\omega - \omega_s$ represents the transient speed of the synchronous machine.
Basically, the SMIB model (\ref{SMIB:ODEs}) adopts the second-order model, which is a reduced model of the fourth-order model described in \eqref{gen:dynamics_gen}, for a single generator that is connected to a very strong grid (infinite bus) through a lossless transmission line without shunt capacitance. 
The parameters $K_1$, $K_2$, and $K_3$ are:
\begin{equation*}
    K_1 = \frac{\omega_s}{2C}T^0_M, K_2 = \frac{\omega_s}{2C}\frac{E_cV}{X}, K_3 = \frac{\omega_s}{2C}D,
\end{equation*}
where $E_c$ represents the magnitude of the machine's internal voltage, $V$ represents the magnitude of the infinite bus voltage, $X$ is equal to the sum of the machine's internal reactance and the lossless line connecting to the infinite bus, which has an angle of zero, and $T^0_M$ represents the constant mechanical torque.
These quantities are in normal per-unit notation, while $C$ and $D$ are the machine inertial constant and the damping constant, respectively, measured in the SI system.
The setting parameters are based on \cite{sauer2017power} (see page 210).

\begin{figure}[t!]
\centering
\begin{subfigure}{0.464\textwidth}
    \begin{subfigure}{0.48\textwidth}
    \includegraphics[width=\textwidth]{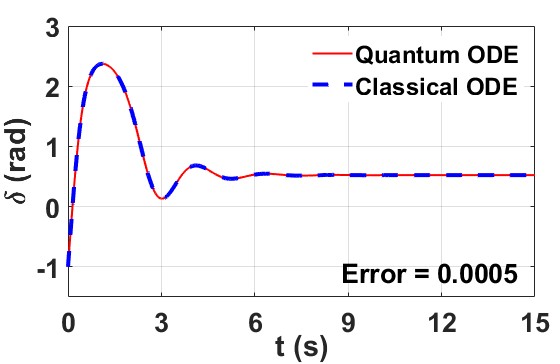}
\end{subfigure}
\begin{subfigure}{0.48\textwidth}
    \includegraphics[width=\textwidth]{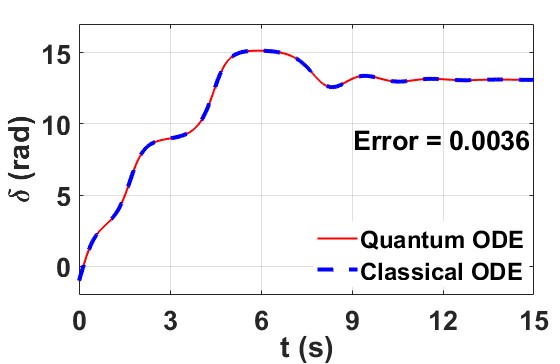}
\end{subfigure}
\end{subfigure}
\begin{subfigure}{0.464\textwidth}
    \begin{subfigure}{0.48\textwidth}
    \includegraphics[width=\textwidth]{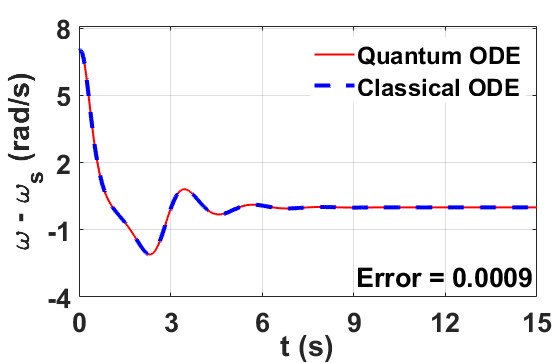}
\end{subfigure}
\begin{subfigure}{0.48\textwidth}
    \includegraphics[width=\textwidth]{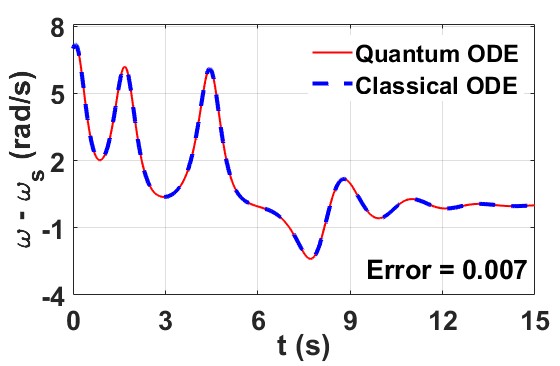}
\end{subfigure}
\end{subfigure}
\begin{subfigure}{0.464\textwidth}
    \begin{subfigure}{0.48\textwidth}
    \includegraphics[width=\textwidth]{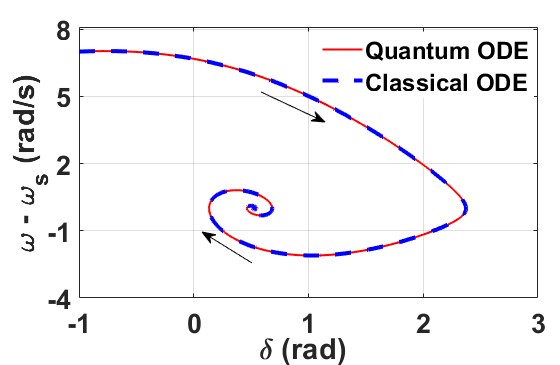}
    \caption{Normal operation}
    \label{fig:normal_operation}
\end{subfigure}
\begin{subfigure}{0.48\textwidth}
    \includegraphics[width=\textwidth]{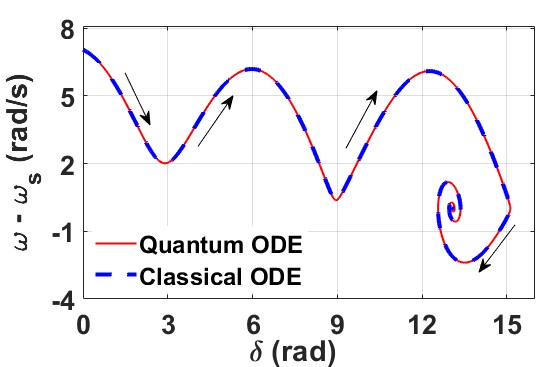}
    \caption{Pole slipping}
    \label{fig:pole_slipping}
\end{subfigure}
\end{subfigure}
\caption{SMIB system's results.}
\label{SMIB}

\end{figure}

We observe the SMIB system in two scenarios: (i) normal operation with $K_1 = 5$, $K_2 = 10$, and $K_3 = 1.7$, and (ii) operation with a special condition, namely pole slipping by changing the parameters $K_3$ to 1.3.
The initial angle is $\delta(0) = - 1rad$, and the initial transient speed is $\omega(0) - \omega_s = 7rad/s$.
The system observation derived from the initial values is shown in Figure \ref{SMIB}.
In both scenarios, the simulation gap between quantum computing and classical methods is very small, with almost zero errors.
In normal operation, the errors between the two methods are 0.0005 and 0.0009 for rotor angle and speed transient, respectively.
When the system is in a pole slipping, the errors increase to 0.0036 and 0.007 because the system takes more time to converge.
In the end, the system reaches a stable state with a phase angle of around $13rad$, which is equal to the phase angle in normal operation ($\approx 0.5rad$) plus two cycles.
The obtained results, including quantum computing and classical methods, are in accordance with the ones in \cite{sauer2017power} (Chapter 5.8).

\subsection{WSCC three-machine nine-bus system}
\begin{figure}[t!]
    \centering
    \includegraphics[width = 0.4\textwidth]{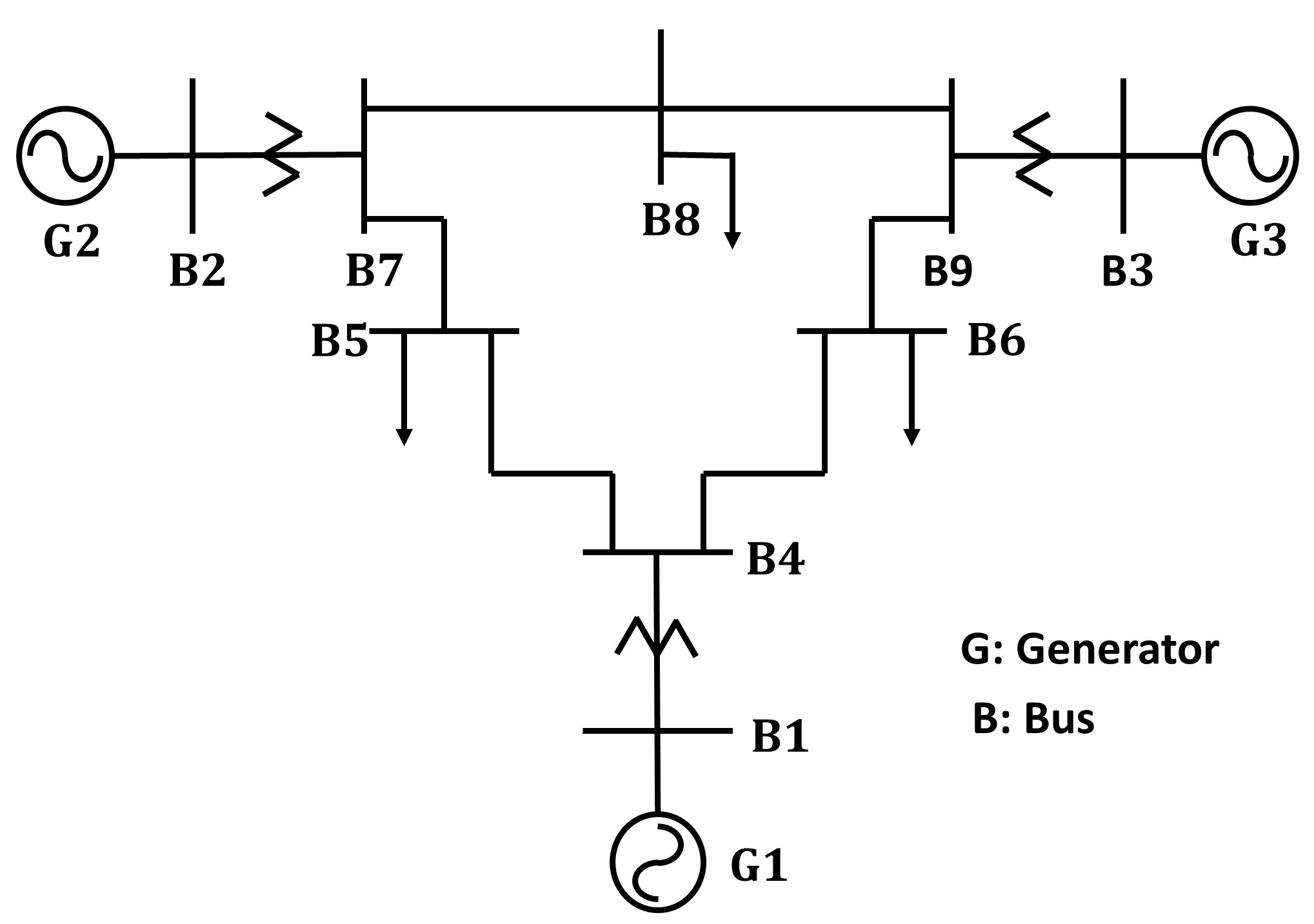}
    \vspace{-0.3em}
    \caption{WSCC three-machine nine-bus system \cite{sauer2017power} (page 143).}
    \label{fig:three}
\end{figure}

The WSCC three-machine nine-bus system depicted in Figure \ref{fig:three} is a basic representation of the Western System Coordinating Council as an equivalent system with three generators and nine buses.
Its parameters can be found in \cite{sauer2017power} (Chapter 7).
We consider two dynamic models: the internal node and the generic DAEs.

\subsubsection{Internal-Node model}
\label{internal}
We have implemented the Internal-Node model, i.e., formulation (\ref{internal_model}), for the WSCC test case, which is a set of ODEs with $12$ state variables. 
The system is assumed to be operating normally until demand disturbances at $t=5$s.
We evaluate the system dynamics in two scenarios:
\begin{itemize}[leftmargin = 15pt]
    \item \textbf{ Decreasing demand:} demands at Bus 6 and Bus 8 decrease by $0.1 + j0.1$pu, and $0.1 + 0.1$pu, respectively.
\item \textbf{ Increasing demand:} demands at Bus 5 and Bus 6 increase by $0.1 + j0.1$pu, and  ${0.1 + j0.1}$pu, respectively.
\end{itemize}

\begin{figure}[t!]
\centering
\begin{subfigure}{0.464\textwidth}
    \begin{subfigure}{0.48\textwidth}
    \includegraphics[width=\textwidth]{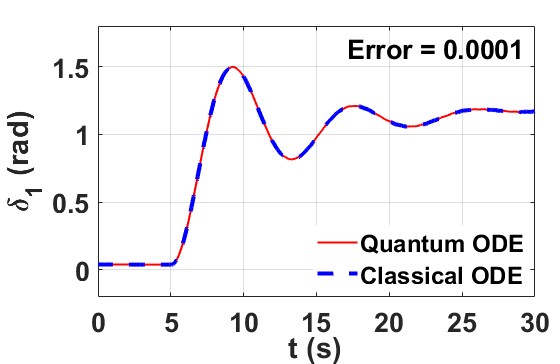}
\end{subfigure}
\begin{subfigure}{0.48\textwidth}
    \includegraphics[width=\textwidth]{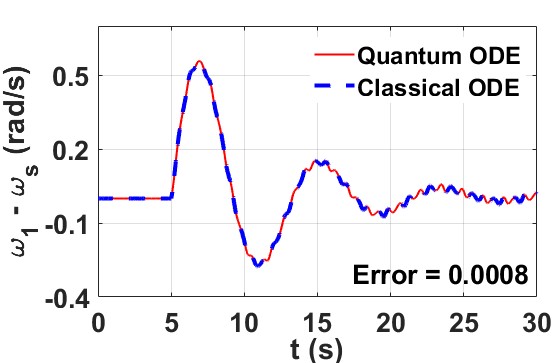}
\end{subfigure}
\vspace{-0.5em}
\caption{Generator 1}
\end{subfigure}
\begin{subfigure}{0.464\textwidth}
    \begin{subfigure}{0.48\textwidth}
    \includegraphics[width=\textwidth]{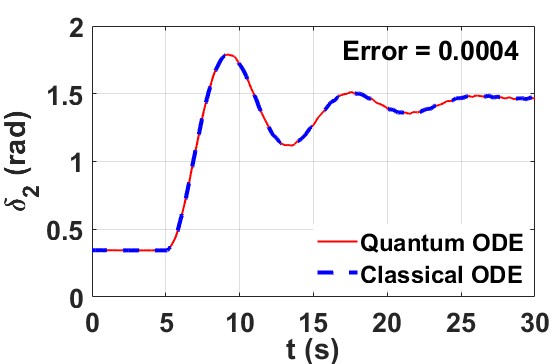}
\end{subfigure}
\begin{subfigure}{0.48\textwidth}
    \includegraphics[width=\textwidth]{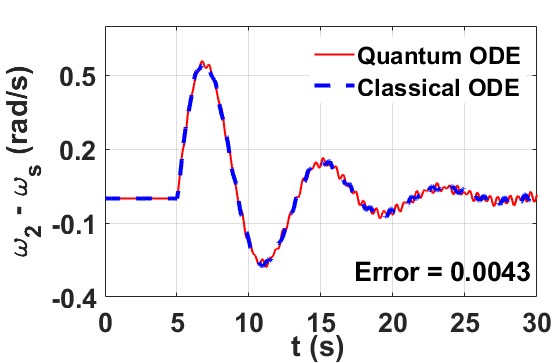}
\end{subfigure}
\vspace{-0.5em}
\caption{Generator 2}
\end{subfigure}
\begin{subfigure}{0.464\textwidth}
    \begin{subfigure}{0.48\textwidth}
    \includegraphics[width=\textwidth]{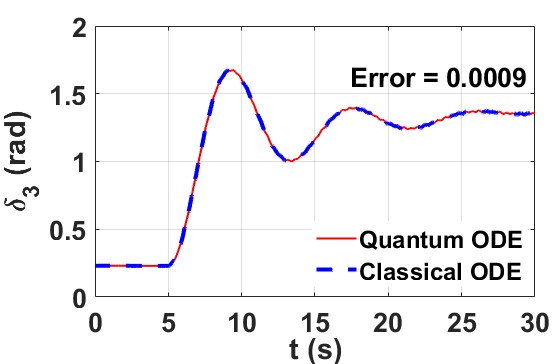}
\end{subfigure}
\begin{subfigure}{0.48\textwidth}
    \includegraphics[width=\textwidth]{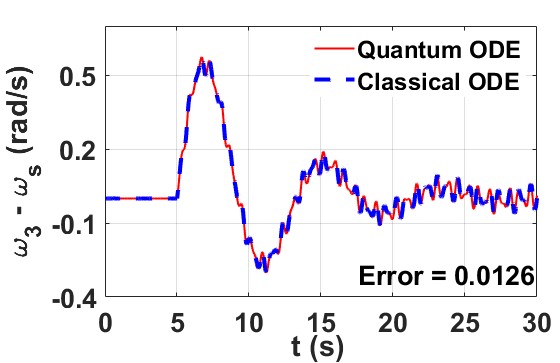}
\end{subfigure}
\vspace{-0.5em}
\caption{Generator 3}
\end{subfigure}
\caption{Internal-Node model with decreasing loads.}
\label{inter_down}
\end{figure}

Figure \ref{inter_down} shows the changes in variables of three machines (phase angle, $\delta_i$, and transient speed $\omega_i - \omega_s$ ($i = 1, 2, 3$) corresponding to generator G1, G2, and G3, respectively).
As the demand suddenly decreases at $t=5$s, the electric power generated from generators is higher than the total load consumption, causing increases in the phase angles and speed of generators.
The system oscillates before converging to a new state, in which all generators' transient speeds converge to zero, thanks to the operation of the governor system.
Phase angles reach new values higher than the initial points.

\begin{figure}[http]
\centering
\begin{subfigure}{0.464\textwidth}
    \begin{subfigure}{0.48\textwidth}
    \includegraphics[width=\textwidth]{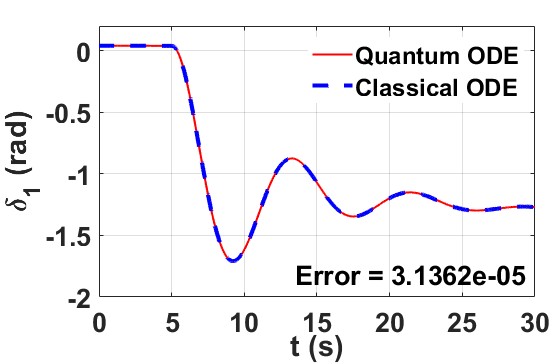}
\end{subfigure}
\begin{subfigure}{0.48\textwidth}
    \includegraphics[width=\textwidth]{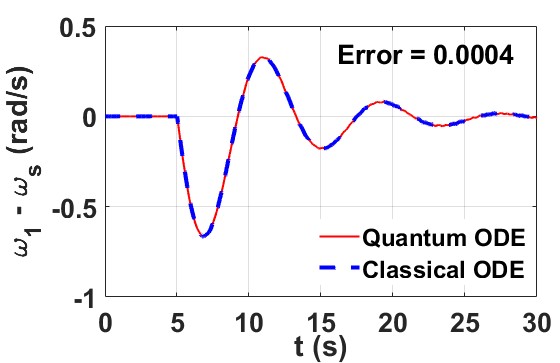}
\end{subfigure}
\vspace{-0.5em}
\caption{Generator 1}
\end{subfigure}
\begin{subfigure}{0.464\textwidth}
    \begin{subfigure}{0.48\textwidth}
    \includegraphics[width=\textwidth]{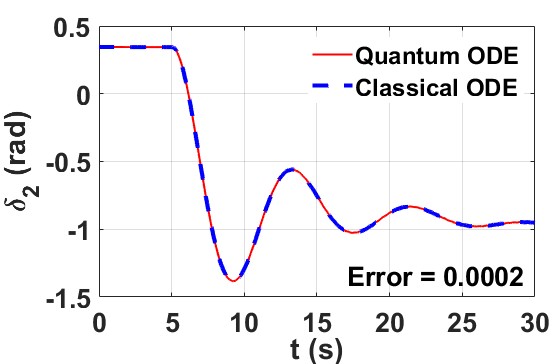}
\end{subfigure}
\begin{subfigure}{0.48\textwidth}
    \includegraphics[width=\textwidth]{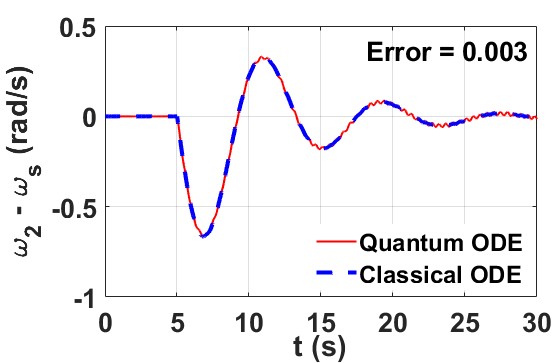}
\end{subfigure}
\vspace{-0.5em}
\caption{Generator 2}
\end{subfigure}
\begin{subfigure}{0.464\textwidth}
    \begin{subfigure}{0.48\textwidth}
    \includegraphics[width=\textwidth]{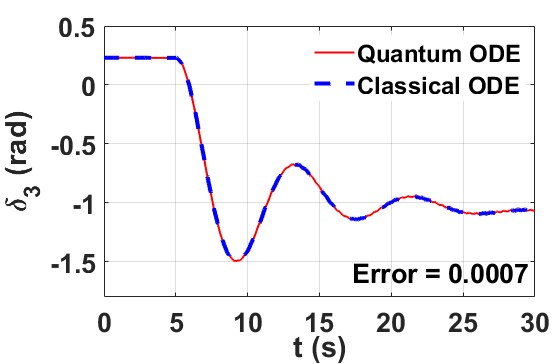}
\end{subfigure}
\begin{subfigure}{0.48\textwidth}
    \includegraphics[width=\textwidth]{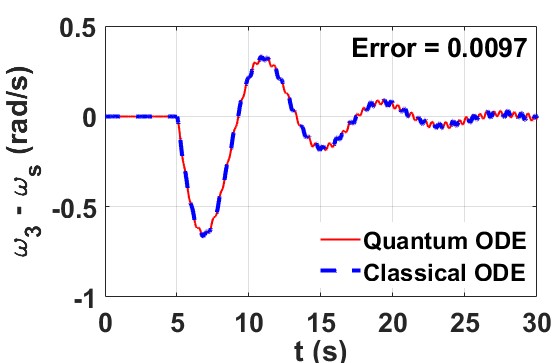}
\end{subfigure}
\vspace{-0.5em}
\caption{Generator 3}
\end{subfigure}
\caption{Internal-Node model with increasing loads.}
\label{inter_up}

\end{figure}

Figure \ref{inter_up} reveals a reverse observation when demand abruptly rises at $t=5$s, leading to insufficient power in the system.
As a result, both phase angles and speeds of the synchronous machines decrease before being stabilized due to the operations of generators' governors, which ramp up power generated in the grid.   
Eventually, the system reaches a new state after oscillating for a while, but the phase angles' values are lower than their pre-disturbance values.

Overall, the obtained results highlight the effectiveness and correctness of quantum computing in solving higher dimensional ODEs, e.g., 12 state variables of the WECC Internal-Node model instead of 2 variables as in SMIB.

\begin{figure}[t!]
\centering
\begin{subfigure}{0.464\textwidth}
\begin{subfigure}{\textwidth}
    \begin{subfigure}{0.48\textwidth}
    \includegraphics[width=\textwidth]{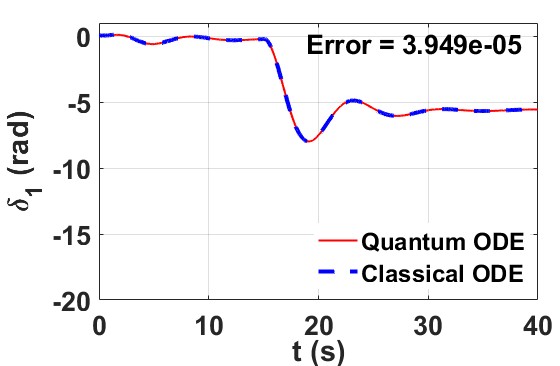}
\end{subfigure}
\begin{subfigure}{0.48\textwidth}
    \includegraphics[width=\textwidth]{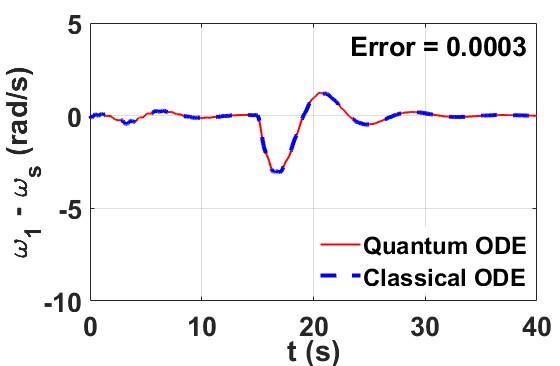}
\end{subfigure}
\end{subfigure}
\begin{subfigure}{\textwidth}
\begin{subfigure}{0.48\textwidth}
    \includegraphics[width=\textwidth]{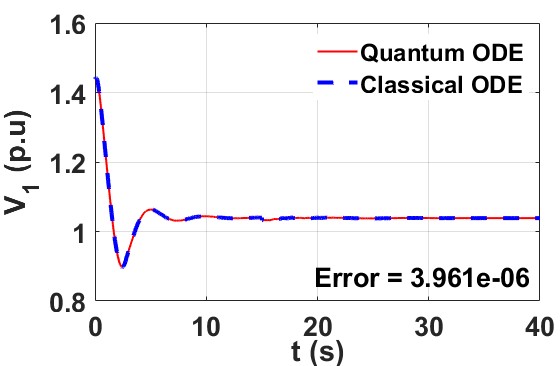}
\end{subfigure}
\begin{subfigure}{0.48\textwidth}
    \includegraphics[width=\textwidth]{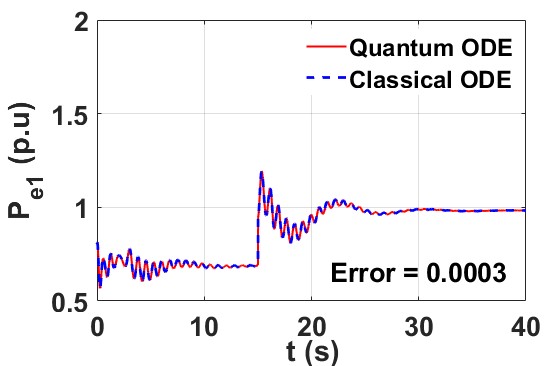}
\end{subfigure}
\end{subfigure}
\vspace{-1.5em}
\caption{Generator 1}
\end{subfigure}
\begin{subfigure}{0.464\textwidth}
\begin{subfigure}{\textwidth}
    \begin{subfigure}{0.48\textwidth}
    \includegraphics[width=\textwidth]{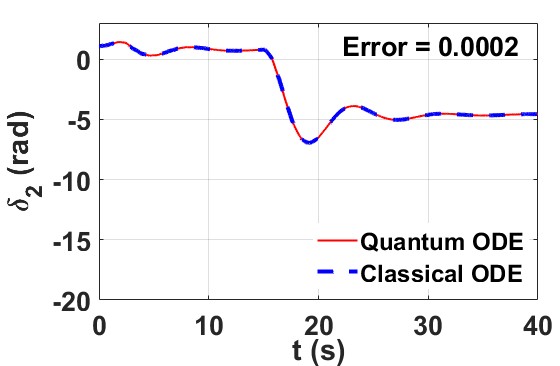}
\end{subfigure}
\begin{subfigure}{0.48\textwidth}
    \includegraphics[width=\textwidth]{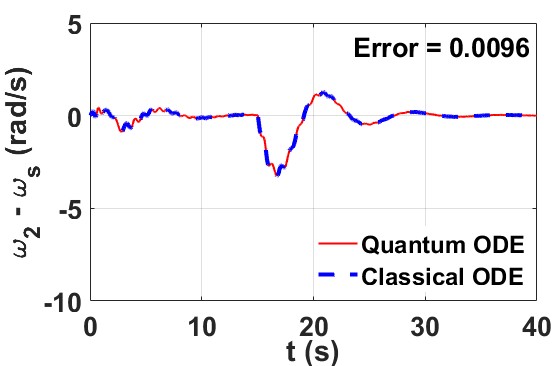}
\end{subfigure}
\end{subfigure}
\begin{subfigure}{\textwidth}
\begin{subfigure}{0.48\textwidth}
    \includegraphics[width=\textwidth]{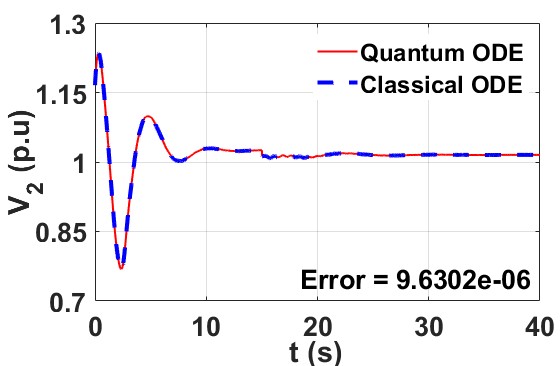}
\end{subfigure}
\begin{subfigure}{0.48\textwidth}
    \includegraphics[width=\textwidth]{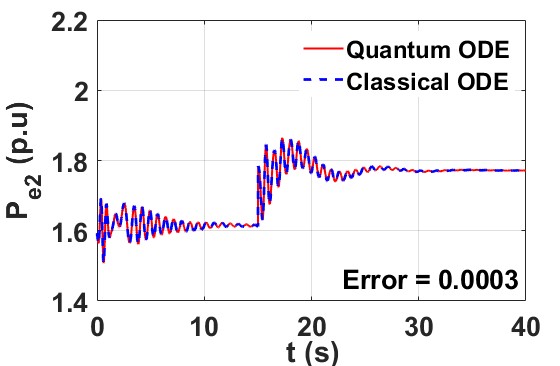}
\end{subfigure}
\end{subfigure}
\vspace{-1.5em}
\caption{Generator 2}
\end{subfigure}
\begin{subfigure}{0.464\textwidth}
\begin{subfigure}{\textwidth}
    \begin{subfigure}{0.48\textwidth}
    \includegraphics[width=\textwidth]{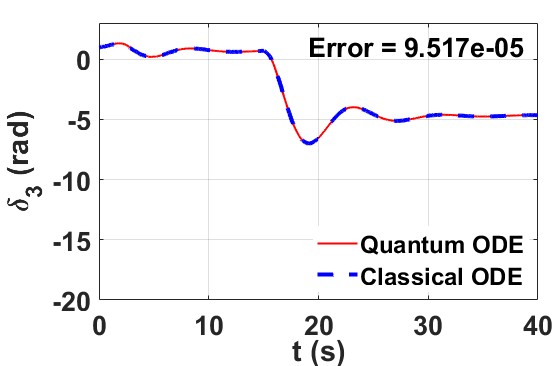}
\end{subfigure}
\begin{subfigure}{0.48\textwidth}
    \includegraphics[width=\textwidth]{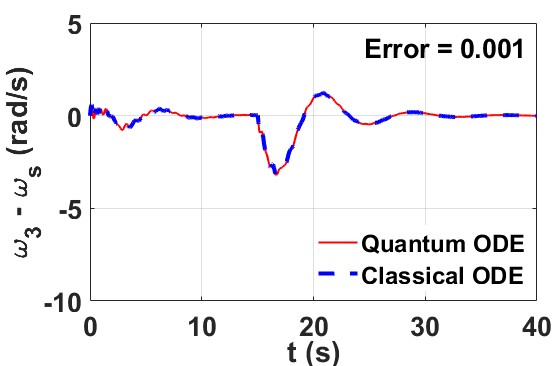}
\end{subfigure}
\end{subfigure}
\begin{subfigure}{\textwidth}
\begin{subfigure}{0.48\textwidth}
    \includegraphics[width=\textwidth]{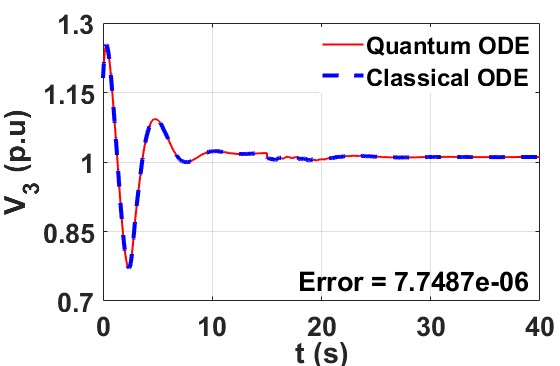}
\end{subfigure}
\begin{subfigure}{0.48\textwidth}
    \includegraphics[width=\textwidth]{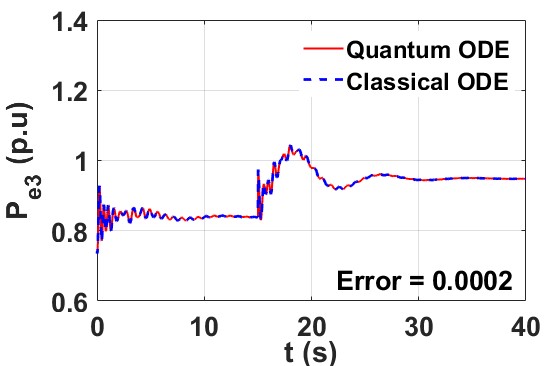}
\end{subfigure}
\end{subfigure}
\vspace{-1.5em}
\caption{Generator 3}
\end{subfigure}
\caption{The generic DAE model with small load changes.}
\label{TM9B_small}
\end{figure}

\subsubsection{Generic DAE model}
\label{Generic}
The generic DAE model as represented in equation (\ref{compact:form}) of the WSCC three-machine nine-bus system consists of 27 state variables and 24 algebraic variables.
By using ModelingToolkit with the Pantelides algorithm, we can convert the DAE model into ODEs (\ref{standardODEs}) with 45 state variables.
Thus, the obtained ODEs are more complicated than the ODEs of the Internal-Node model.  
At the time $t = 0$s, we set the internal voltage on the $q$-axis of generator G1, ${E'}_{q,1}(0)$, to be 0.45pu above the normal level and its angle $\delta_{1}(0)$ to be 0.01pu departed from the equilibrium.   
This enables us to simulate the system behavior with the initial over-voltage of G1.  
We consider two scenarios of load disturbances at $t = 15$s as follows:
\begin{itemize}[leftmargin = 10pt]
    \item \textbf{ Small disturbance:} demands at Bus 5, Bus 6, and Bus 8 increase by ${0.2} {+} {j}{0.05}$pu, $0.15$pu, and  $0.2$pu, respectively.
    
\item \textbf{ Large disturbance:} demands at Bus 5, Bus 6, and Bus 8 increase by ${0.5} {+} {j}{0.25}$pu,  ${0.3}{+}{j}{0.05}$pu, and  ${0.4}{+}{j}$0.05pu respectively.
\end{itemize}

\begin{figure}[t!]
\centering
\begin{subfigure}{0.464\textwidth}
\begin{subfigure}{\textwidth}
    \begin{subfigure}{0.48\textwidth}
    \includegraphics[width=\textwidth]{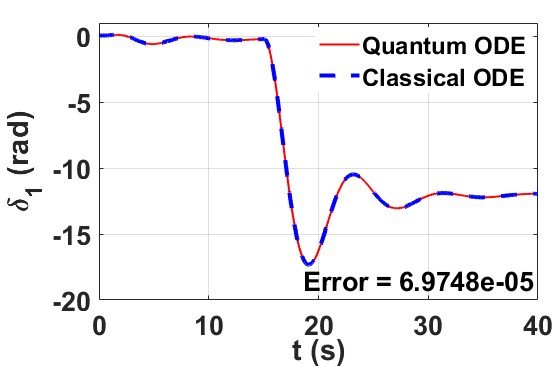}
\end{subfigure}
\begin{subfigure}{0.48\textwidth}
    \includegraphics[width=\textwidth]{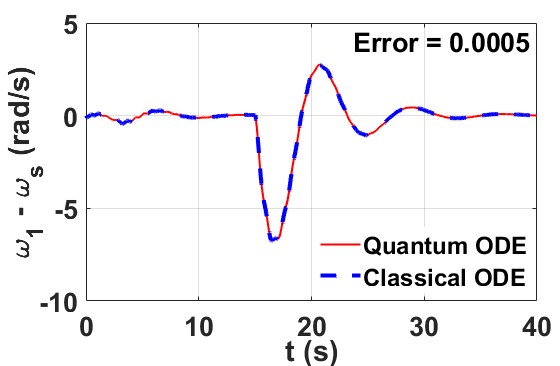}
\end{subfigure}
\end{subfigure}
\begin{subfigure}{\textwidth}
\begin{subfigure}{0.48\textwidth}
    \includegraphics[width=\textwidth]{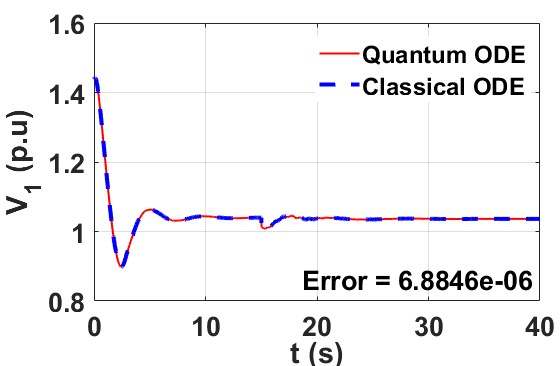}
\end{subfigure}
\begin{subfigure}{0.48\textwidth}
    \includegraphics[width=\textwidth]{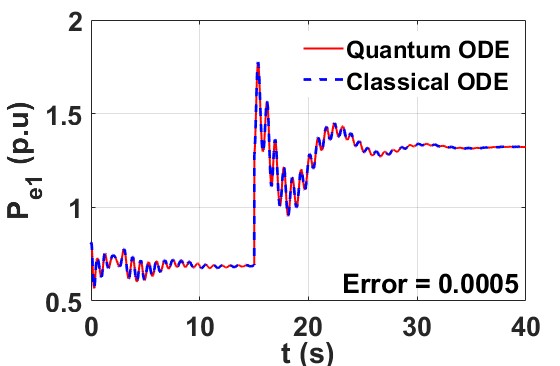}
\end{subfigure}
\end{subfigure}
\vspace{-1.5em}
\caption{Generator 1}
\end{subfigure}
\begin{subfigure}{0.464\textwidth}
\begin{subfigure}{\textwidth}
    \begin{subfigure}{0.48\textwidth}
    \includegraphics[width=\textwidth]{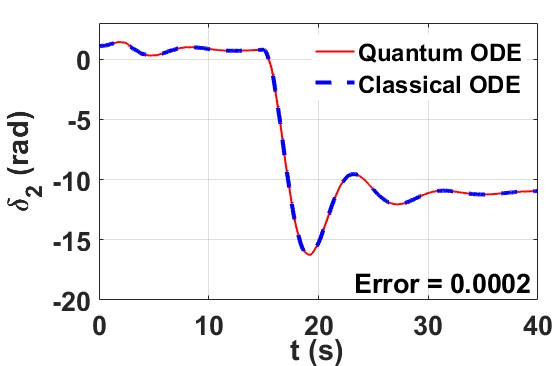}
\end{subfigure}
\begin{subfigure}{0.48\textwidth}
    \includegraphics[width=\textwidth]{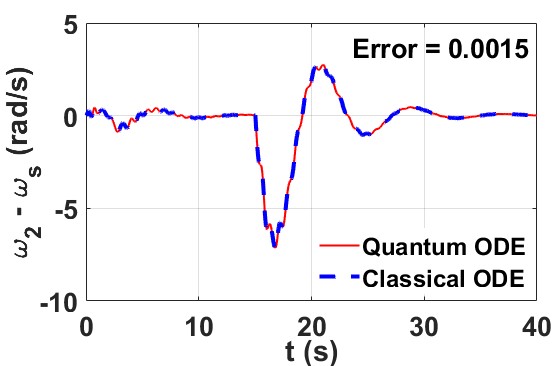}
\end{subfigure}
\end{subfigure}
\begin{subfigure}{\textwidth}
\begin{subfigure}{0.48\textwidth}
    \includegraphics[width=\textwidth]{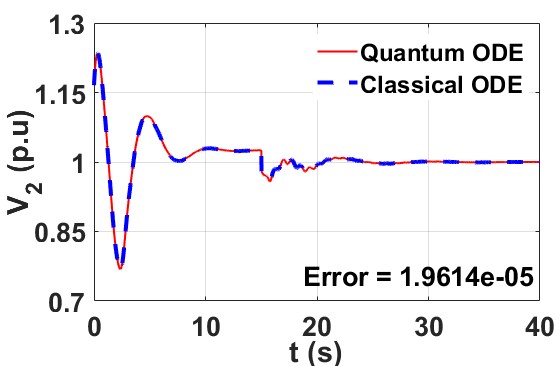}
\end{subfigure}
\begin{subfigure}{0.48\textwidth}
    \includegraphics[width=\textwidth]{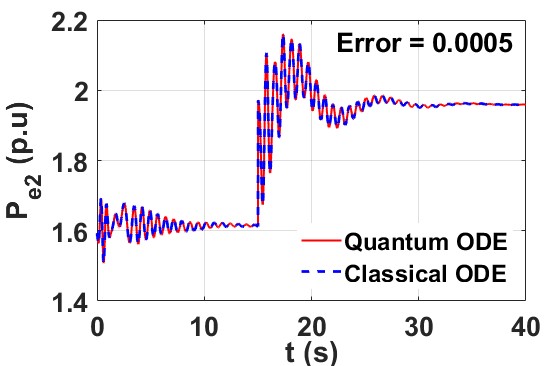}
\end{subfigure}
\end{subfigure}
\vspace{-1.5em}
\caption{Generator 2}
\end{subfigure}
\begin{subfigure}{0.464\textwidth}
\begin{subfigure}{\textwidth}
    \begin{subfigure}{0.48\textwidth}
    \includegraphics[width=\textwidth]{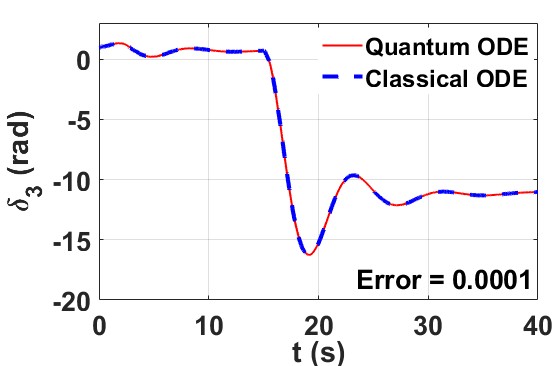}
\end{subfigure}
\begin{subfigure}{0.48\textwidth}
    \includegraphics[width=\textwidth]{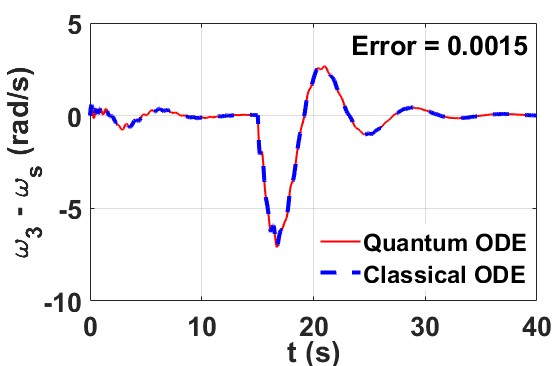}
\end{subfigure}
\end{subfigure}
\begin{subfigure}{\textwidth}
\begin{subfigure}{0.48\textwidth}
    \includegraphics[width=\textwidth]{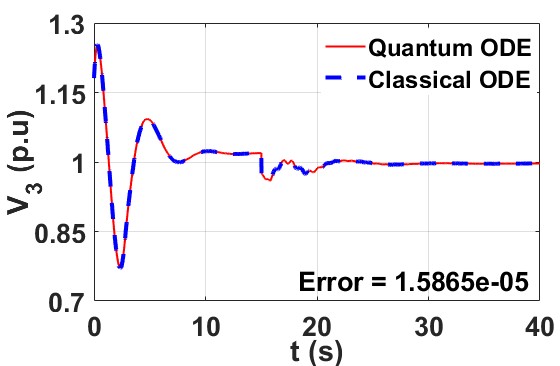}
\end{subfigure}
\begin{subfigure}{0.48\textwidth}
    \includegraphics[width=\textwidth]{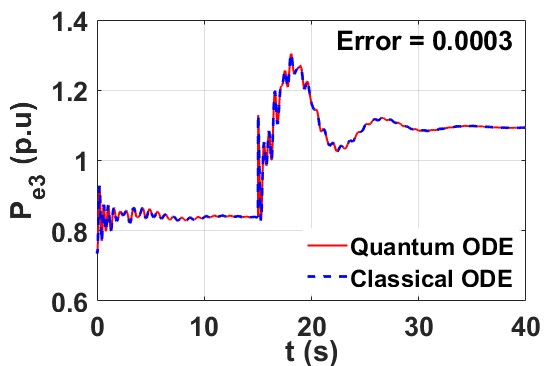}
\end{subfigure}
\end{subfigure}
\vspace{-1.5em}
\caption{Generator 3}
\end{subfigure}
\vspace{-0.5 em}
\caption{The generic DAE model with large load changes.}
\label{TM9B_large}
\vspace{-4 mm}
\end{figure}

Figure \ref{TM9B_small} displays the changes of variables, i.e., phase angle, $\delta_i$, transient speed, $\omega_i - \omega_s$, the generator's voltage, $V_i$, and active power output $P_{ei}$ ($i=1,2,3$) for generators G1, G2 and G3.
Note the active power output is calculated as: 
\beqn
   P_{ei} = E'_{di}I_{di}+E'_{qi}I_{qi}+(X'_{qi}-X'_{di})I_{di}I_{qi}
\eeqn
Between $t=0$s and $t=15$s, all generators' variables undergo a brief oscillation from their initial values before eventually returning to their equilibrium values.
When the load changes at $t=15$s, all generators' outputs increase to balance the demand while their phase angles, speeds, and voltages decrease.
During both periods ( $0\leq t <15$s and $t\geq 15$s), the system stabilizes thanks to the operations of the exciters and governors for stabilizing generators' voltages and speeds accordingly. 
While the transient speed and voltages of generators return to the equilibrium points around zero and 1.0pu, respectively, phase angles and active electric power outputs reach new equilibrium points after the load demand changes.

Figure \ref{TM9B_large} shows similar results with a higher magnitude of oscillation of all variables due to a higher change of loads.
The system also moves to a new equilibrium of operating points in which power generated by generators can match the new demands thanks to the operations of their governors. 
Overall, the simulation results in both cases illustrate the potential of quantum computing algorithms in solving generic DAEs modeling power system dynamics as well as the effectiveness of scientific machine learning tool-kits, e.g., the Julia-based ModelingToolkit, for symbolically representing and transforming system dynamics. 

\section{CONCLUSION}
\label{conclu}
This paper presents a quantum computing approach for solving DAEs in power system dynamics.
We review the mathematical models of power system dynamics, particularly the original DAE model in a generic setting and the simplified ODE one in the internal node assumption.   
The original DAEs are first converted into a set of nonlinear ODEs using the index reduction method.
Using amplitude encoding, the state vector can be encoded into a quantum computer that requires the number of qubits as a logarithm of the number of state variables.
To exploit the quantum advantages in complex linear algebra, nonlinear ODEs are polynomially approximated as a set of polynomial functions of state variables, which can be encoded in a quantum computer by probability amplitudes of tensors of quantum states and Hamiltonian simulation. 
The linear update in traditional ODE solvers, e.g., the Forward-Euler equation, can be conducted by quantum-based linear solvers, such as the HHL algorithm.
We demonstrate the potential of the quantum computing approach on two power network cases, particularly the single-machine infinite bus system and the Western System Coordinating Council three-machine nine-bus system.
Our numerical results conducted on different dynamic models of these test cases demonstrate the potential of quantum computing in analyzing power system dynamics with high accuracy.
Additionally, our study also illustrates the use of scientific machine learning tools, particularly symbolic programming, to facilitate the use of complex computing concepts, e.g., Taylor expansion, DAEs/ODEs transformation, quantum circuits, and quantum operators, in the field of power engineering.

\section{ACKNOWLEDGEMENT}
This research is supported in part by the Alfred P. Sloan Foundation under grant \#10358, in part by the North Carolina Agricultural and Technical State University's Intel Foundation Gift, and in part by the DOE Sandia National Laboratories under contract \#281247.

\bibliographystyle{IEEEtran}
\renewcommand{\refname}{REFERENCES}
\bibliography{paper_references}

\vspace{-\topsep}
\section{APPENDICES}\label{apend}

\subsection{Ket and Bra notation}
\label{ket_bra}

In quantum mechanics, the convenient way to represent the spin state is using Dirac notation \cite{lapierre2021introduction}, which puts a description of the state in brackets.
The ``$\ket{}$'' is called a ``ket'', representing the column vector, e.g., the quantum state of a qubit $ \ket{\psi}  = \alpha \ket{0} + \beta \ket{1} \rightarrow     \begin{bmatrix}
        \alpha \\ \beta
    \end{bmatrix} $.
The Dirac bra ``$\bra{\psi}$'' represent the ``dual vector'' of the Dirac ket $\ket{\psi}$:
\begin{equation*}
     \bra{\psi}  = \ket{\psi}^\dag = ((\ket{\psi})^\top)^*= \alpha^*\bra{0} + \beta^*\bra{1} \rightarrow [\alpha^* ~ \beta^*]
\end{equation*}
where ``$\top$'' is used to take the transpose of the vector, and ``$*$'' is used to take the complex conjugate of each vector element.
The combination of these two actions is called the ``conjugate transpose'', the ``adjoint'' or ``Hermitian conjugate'', denoted as ``$\dag$''.

\subsection{Explain the quantum steady-state  with Von-Neuman Hamiltonian trick in (\ref{steady_state})} \label{Steady-state}
\bse
We now explain how the Hamiltonian operator $H$  of the Von-Neumann measurement prescription leads to quantum steady state $\ket{\Psi}$ in (\ref{steady_state}). 
Using $H$ defined in (\ref{halminton}) and initial state $\ket{z} \ket{z} \ket{0}$, the steady quantum state is $\ket{\Psi} = e^{-i \epsilon H} \ket{z} \ket{z} \ket{0} = \sum\limits_{j=0}^\infty \frac{ (-i \epsilon H)^j}{j!} \ket{z} \ket{z} \ket{0}$ following the discussion in section \ref{time_evolution}.    
The element of the Taylor expansion with $j=0$ is:
\begin{equation}
\frac{(-i \epsilon H)^0}{0!}\ket{z}\ket{z}\ket{0} = \ket{z}\ket{z}\ket{0}, \label{j=0}
\end{equation}
which is the first term in (\ref{steady_state}), and the element of the Taylor expansion with $j=1$ is:
\begin{align}
    &\frac{(-i \epsilon H)^1}{1!} \ket{z}\ket{z}\ket{0} \label{j=1} \\
    &= \underbrace{(\epsilon A \otimes \ket{1}_P \bra{0})\ket{z}\ket{z}\ket{0}}_{F_1} + \underbrace{( -\epsilon A^\dag \otimes \ket{0}_P \bra{1})\ket{z}\ket{z}\ket{0}}_{F_2}, \nonumber
\end{align}
which will be proven to be the second term in (\ref{steady_state}) as follows. 
The term $F_1$ is originally expressed as follows:
\begin{equation}
    F_1 = (\epsilon A \otimes \ket{1}_P \bra{0})(\ket{z}\ket{z} \otimes \ket{0}).
\end{equation}
By utilizing the ``no name'' property of the tensor product in Table 9.1 of reference \cite{lapierre2021introduction}, i.e., $ (A \otimes B)(C \otimes D) = (AC)\otimes(BD) $, we can rewrite $F_1$ as:
\begin{equation}
    F_1 = (\epsilon A \ket{z}\ket{z}) \otimes (\ket{1}_P \bra{0}\ket{0}).
\end{equation}
Then, following the inner product principle in \cite{lapierre2021introduction}, Chapter 4.6, page 62, and note that $\braket{0|0} = 1$, we have:
\begin{align}
    F_1 & = (\epsilon A \ket{z}\ket{z}) \otimes (\ket{1}_P \braket{0|0}) \\ & = \epsilon A \ket{z}\ket{z} \otimes \ket{1} = \epsilon A \ket{z}\ket{z}\ket{1}, \nonumber
\end{align}
Similarly, applying the inner product principle, we have $F_2 = 0$ due to $\braket{1|0} = 0$.
Thus, equation (\ref{j=1}) becomes the second term in (\ref{steady_state}):
\begin{equation}
    \frac{(-i \epsilon H)^1}{1!} \ket{z}\ket{z}\ket{0} = \epsilon A \ket{z}\ket{z}\ket{1}.
\end{equation}

\ese
\subsection{HHL algorithm} \label{HHL_algorithm}

\begin{figure}[t!]
    \centering
    \includegraphics[width = 0.95\linewidth]{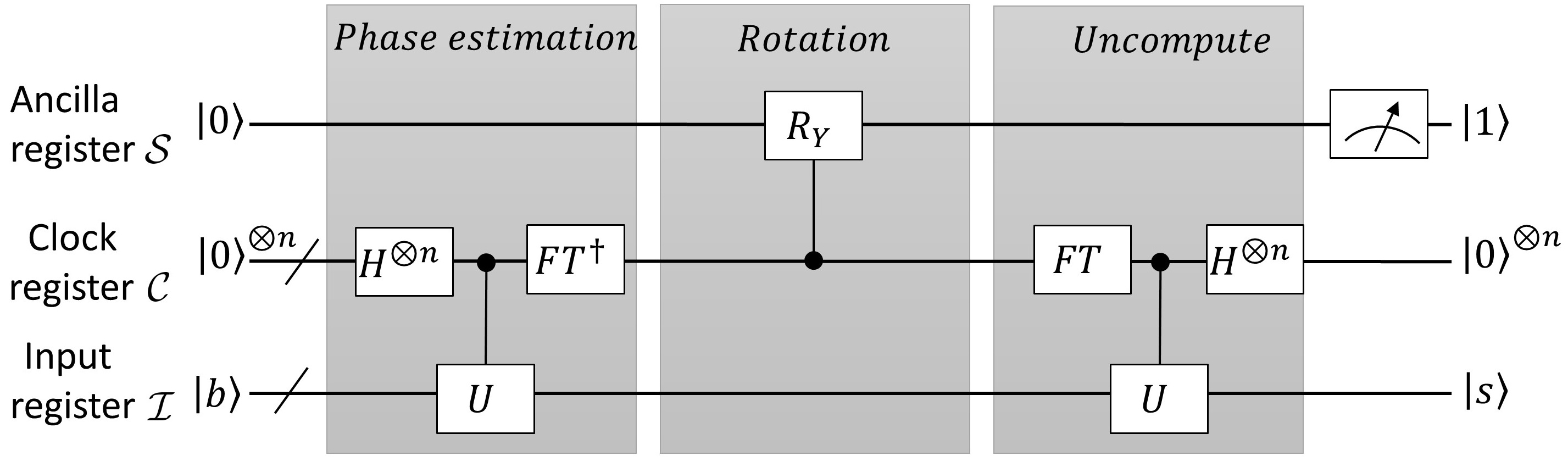}
    \caption{Overview of the HHL algorithm circuit ($n=\log_2 N$).}
    \label{fig:HHL_circuit}
\end{figure}

\bse
Given a linear system with a Hermitian $N \times N$ matrix $M$ and a normalized column vector $b$, we want to find a vector $s$ satisfying $M s = b$ in a quantum computer.
Based on linear algebra,
we can express $M$ via the sum of the outer products of its eigenvectors scaled by its eigenvalues as $
M = \sum_{j=1}^N \lambda_j \ket{u_j}\bra{u_j} $ and its inverse is $ M^{-1} = \sum_{j=1}^N \lambda_j^{-1} \ket{u_j}\bra{u_j}$ where $\lambda_j$ denotes the eigenvalue and $\ket{u_j}$ denotes the eigenvectors of $M$, ($j$=1,$\ldots$,$N$).
Since $M$ is invertible and Hermitian, it must have an orthogonal basis of eigenvectors $\ket{u_j}$, and vector $b$ can be expressed as $\ket{b} = \sum_{j=1}^N b_j\ket{u_j}$ \cite{harrow2009quantum}.
Therefore, the solution of $M \ket{s} = \ket{b}$ can be written as:
\beqn
{\footnotesize	
    \ket{s} = M^{-1} \ket{b} = \sum_{j=1}^N \text{\boldmath  $\frac{b_j}{\lambda_j}$ \unboldmath} \ket{u_j}, \label{eq:solution} 
\text{i.e.,}~ s =\Big[ \text{\boldmath  $\frac{b_j}{\lambda_j}$ \unboldmath}, \forall j =1,\ldots,N \Big]^\top } 
\eeqn
If $M$ is not Hermitian, we can apply the reduction of $M$ to get Hermitian $\widehat{M}$, then the problem $M \ket{s} = \ket{b}$ is equivalent to:
\begin{equation*}
    \underbrace{\begin{bmatrix}
        0 & M \\
        M^\dag & 0
    \end{bmatrix}}_{\widehat{M}} \underbrace{\begin{bmatrix}
        0 \\
        s
    \end{bmatrix}}_{\ket{\widehat{s}}} = \underbrace{\begin{bmatrix}
        b \\
        0
    \end{bmatrix}}_{\ket{\widehat{b}}}.
\end{equation*}
The rest of this section presents the algorithm for the Hermitian matrix. 
The HHL algorithm depicted in Figure \ref{fig:HHL_circuit} aims to obtain a quantum state $\ket{s}$ that satisfies (\ref{eq:solution}). 
It has three steps: phase estimation, eigenvalue inversion rotation, and inverse phase estimation, and employs three quantum registers: the ancilla register $\mathcal{S}$ storing auxiliary qubits, the clock registers $\mathcal{C}$ storing a binary representation of the eigenvalues of matrix $M$, and the input register $\mathcal{I}$ storing the vector solution of the system of linear equations when the measurement of the contents of the ancilla quantum register is ``1''.

Applying quantum phase estimation with the unitary operator $U = e^{iMt}$, the register expressed in the eigenbasis of $M$ becomes:
\begin{equation}
    \sum_{j=1}^N b_j \ket{\lambda_j}_\mathcal{S} \ket{u_j}_\mathcal{I}.
\end{equation}

\noindent
Applying rotation of ancilla qubit conditioned on each eigenvalue $\ket{\lambda_j}$ of $M$, we get a normalized state of the form:
\begin{equation}
  \sum_{j = 1}^N b_j \ket{\lambda_j}_\mathcal{S} \ket{u_j}_\mathcal{I} \left( \sqrt{1-\frac{c^2}{\lambda_j^2}} \ket{0} + \frac{c}{\lambda_j} \ket{1} \right),  \label{eq: control_rotation}
\end{equation}
where $c$ is the scaling factor used to guarantee that all quantum states are normalized.  
Then, we perform inverse quantum phase estimation, which sets the clock register $\mathcal{C}$ back to 0's and leaves the remaining states as:

\begin{equation}
  \sum_{j = 1}^N b_j \ket{0}_\mathcal{S} \ket{u_j}_\mathcal{I} \left( \sqrt{1-\frac{c^2}{\lambda_j^2}} \ket{0} + \frac{c}{\lambda_j} \ket{1} \right). \label{eq: control_reverse}
\end{equation}
We observe that if the ancilla qubit is 1, the state of the registers is a normalized equivalent of the solution vector $s$ (note that  $\|s\| = \sqrt{\sum_{j=1}^N |b_j|^2/|\lambda_j|^2}$).
Thus, measuring the ancilla qubit until getting ``1'' and the new state is:  

\begin{equation}
    \sqrt{\frac{1}{\sum_{j=1}^N |b_j|^2/|\lambda_j|^2}}  \sum_{j = 1}^N \text{\boldmath $\frac{b_j}{\lambda_j}$ \unboldmath} \ket{0}_\mathcal{S} \ket{u_j}_\mathcal{I}, \label{eq: final_state}
\end{equation}
which is component-wise proportional to the solution vector $\ket{s}$, i.e., the values of $\dfrac{b_j}{\lambda_j}$ can be extracted from quantum states of register $\mathcal{I}$.

\ese

\end{document}